\documentclass[aps,prd,nofootinbib,amsmath,amssymb,showpacs,longbibliography,singlecolumn,superscriptaddress,notitlepage]{revtex4-1}

\usepackage{epsfig}
\usepackage{bm}
\usepackage{amssymb}
\usepackage{amsmath}
\usepackage{dsfont}
\usepackage{color}
\usepackage{subfigure}
\usepackage{dcolumn}
\usepackage{mathtools}
\usepackage{enumerate}
\usepackage[utf8]{inputenc}
\usepackage{soul}

\usepackage{amsmath}
\usepackage{amssymb}
\usepackage{amsthm}
\usepackage{amscd, amsfonts, mathrsfs}
\usepackage{cases}
\usepackage{verbatim} 
\usepackage{epsf}
\usepackage[bookmarksnumbered,pdfpagelabels=true,plainpages=false,colorlinks=true,
            linkcolor=black,citecolor=black,urlcolor=black]{hyperref}

\theoremstyle{plain}

\theoremstyle{definition}

\allowdisplaybreaks

\newcommand*{\Scale}[2][4]{\scalebox{#1}{$#2$}}%
\newcommand{\forceds}[1]{\Scale[1.1]{\ensuremath{\mathrlap{{{#1}}}\hspace{1.4pt}#1}}}
\newcommand{\cvec}[1]{\ensuremath{\bm {#1}}}
\newcommand{\cmat}[1]{\ensuremath{\mathds{{#1}}}}
\newcommand{\cmatgreek}[1]{\ensuremath{\forceds{{#1}}}}

\makeatletter
\newcommand{\unitaryi}[2]{\ensuremath{U_{(#1,#2)}}}
\newcommand{\unitarydaggi}[2]{\ensuremath{U_{(#1,#2)}}}
\newcommand{\unitaryinsertfield}[1]{\ensuremath{[#1]}}
\newcommand{\@unitaryii}{\@ifnextchar\bgroup{\unitaryinsertfield}{}}
\newcommand{\unitary}[2]{\unitaryi{#1}{#2}\@unitaryii}
\newcommand{\unitarydagg}[2]{\unitarydaggi{#1}{#2}\@unitaryii}
\makeatother

\newcommand{\tr}{\operatorname{tr}}

\newcommand{\be}{\begin{equation}}
\newcommand{\ee}{\end{equation}}
\newcommand{\bea}{\begin{eqnarray}}
\newcommand{\eea}{\end{eqnarray}}

\newcommand{\bml}{\begin{subequations}}
\newcommand{\eml}{\end{subequations}}

\newcommand{\bbm}{\begin{bmatrix}}
\newcommand{\ebm}{\end{bmatrix}}
\newcommand{\bvm}{\begin{vmatrix}}
\newcommand{\evm}{\end{vmatrix}}

\begin{document}


\title{Relativistic hydrodynamic fluctuations from an effective action: causality, stability, and the information current}
\date{\today}

\author{Nicki Mullins}
\email{nickim2@illinois.edu}
\affiliation{Illinois Center for Advanced Studies of the Universe\\ Department of Physics, 
University of Illinois at Urbana-Champaign, Urbana, IL 61801, USA}

\author{Mauricio Hippert}
\email{hippert@illinois.edu}
\affiliation{Illinois Center for Advanced Studies of the Universe\\ Department of Physics, 
University of Illinois at Urbana-Champaign, Urbana, IL 61801, USA}

\author{Lorenzo Gavassino}
\email{lorenzo.gavassino@vanderbilt.edu}
\affiliation{Department of Mathematics, Vanderbilt University, Nashville, TN, USA}

\author{Jorge Noronha}
\email{jn0508@illinois.edu}
\affiliation{Illinois Center for Advanced Studies of the Universe\\ Department of Physics, University of Illinois at Urbana-Champaign, Urbana, IL 61801, USA}

\begin{abstract}
Causality is necessary for retarded Green's functions to remain retarded in all inertial frames in relativity, which ensures that dissipation of fluctuations is a Lorentz invariant concept. For first-order BDNK theories  with stochastic fluctuations, introduced via the Schwinger-Keldysh formalism, we show that imposing causality and stability leads to correlation functions of hydrodynamic fluctuations that only display the expected physical properties at small frequencies and wavenumber, i.e., within the expected regime of validity of the first-order approach. For second-order theories of Israel and Stewart type, constructed using the information current such that entropy production is always non-negative, a stochastic formulation is presented using the Martin-Siggia-Rose approach where imposing causality and stability leads to correlators with the desired properties. We also show how Green's functions can be determined from such an action. We identify a $\mathds{Z}_2$ symmetry, analogous to the Kubo-Martin-Schwinger symmetry, under which this Martin-Siggia-Rose action is invariant. 
This modified Kubo-Martin-Schwinger symmetry provides a new guide for the effective action formulation of hydrodynamic systems with dynamics not solely governed by conservation laws. Furthermore, this symmetry ensures that the principle of detailed balance is valid in a covariant manner. We employ the new symmetry to further clarify the connection between the Schwinger-Keldysh and Martin-Siggia-Rose approaches, establishing a precise link between these descriptions in second-order theories of  relativistic hydrodynamics. 
Finally, the modified Kubo-Martin-Schwinger symmetry is used to determine the corresponding action describing diffusion in Israel-Stewart theories in a general hydrodynamic frame.  
\end{abstract}

\maketitle

\section{Introduction}

Much of modern theoretical physics is constructed from the framework of effective field theory (EFT); the idea that physical models can be systematically built, order-by-order in some expansion parameter, using  effective degrees of freedom and symmetries \cite{Weinberg:1995mt, Weinberg:1996kr}. A classic example of this approach is fluid dynamics, which describes the late-time, long-wavelength evolution of some set of conserved quantities, such as energy, momentum, and particle (baryon) number \cite{Landau1987Fluid}. In the non-relativistic limit, viscous fluids are traditionally studied using the famous Navier-Stokes equations \cite{Landau1987Fluid}, which are known to describe a wide range of physical phenomena \cite{Alfvn1942ExistenceOE,pope_2000}. 

One key feature present in fluids is the existence of irreversible processes and, hence, dissipation. The fluctuation-dissipation theorem \cite{Callen:1951vq, Kubo:1957mj} dictates that a thermal system with dissipation will also experience fluctuations in its thermodynamic quantities. Therefore, a complete EFT approach for hydrodynamics requires the proper inclusion of these fluctuations. Typically, thermal fluctuations are modeled by including stochastic sources in the hydrodynamic theory, resulting in stochastic partial differential equations \cite{landau_statistical_part_II}. While in principle such equations can be solved directly, much progress has been made by recasting the stochastic dynamics in a path integral form. Using this approach, the fluctuating hydrodynamic system is mathematically similar to a quantum field theory, allowing for powerful field theory techniques to be employed \cite{Baym:1961zz, Schwinger:1960qe, Keldysh:1964ud, Martin:1973zz, Zinn-Justin:572813}. In fact, these techniques have been widely applied to study non-relativistic fluctuating systems \cite{weiss2008quantum}. However, fundamental challenges still abound when it comes to relativistic fluids.

Early attempts to generalize the Navier-Stokes equations to relativity were made by Eckart \cite{PhysRev.58.919} and Landau and Lifshitz \cite{Landau1987Fluid}. While the EFT construction of these theories is in principle similar to that of the Navier-Stokes equations, namely a first-order derivative expansion, they were later found to possess unphysical behavior signaled by causality violation \cite{PichonViscous}. Furthermore, such theories predicted that the number of collective modes in the shear and sound channels depends on the Lorentz frame, signaling that the global equilibrium state is unstable with respect to small perturbations \cite{Hiscock:1985zz}. This is especially problematic for the study of thermal fluctuations in this relativistic system, as it indicates that there can be fluctuations that will rapidly take the system away from equilibrium, never to return.

Historically, these issues of causality and stability were first repaired through the construction of the so-called second-order theories \cite{1967ZPhy..198..329M, MIS-2,MIS-6}. The latter, which we will refer to here as Israel-Stewart theory, is qualitatively different from the first-order theories of Eckart, and Landau and Lifshitz. In fact, in second-order theories, dissipative contributions to the energy-momentum tensor and conserved currents obey their own dynamical equations of motion, which can be derived from a number of different approaches \cite{MIS-6, Baier:2007ix, Denicol:2012cn, MuellerRuggeriBook, JouetallBook,GavassinoGENERIC2022}. Therefore, when compared to first-order approaches, Israel-Stewart theory is said to possess an extended set of variables as it treats dissipative fluxes as legitimate new degrees of freedom in addition to the standard hydrodynamic variables. Israel-Stewart theory is currently the prevalent approach used in numerical studies of relativistic fluids, see e.g. \cite{Romatschke:2017ejr}.  

Recently, it has become clear that first-order theories can \emph{also} be causal and stable if one uses hydrodynamic fields defined in a way different than done by Eckart and Landau-Lifshitz. This recent development, due to Bemfica, Disconzi, Noronha, and Kovtun (BDNK) \cite{BemficaDisconziNoronha, Kovtun:2019hdm, Bemfica:2019knx, Hoult:2020eho,Bemfica:2020zjp}, relies on the fact \cite{MIS-2,MIS-6} that hydrodynamic quantities such as temperature, chemical potential, and fluid velocity are not uniquely defined out of equilibrium. A choice to define these variables always has to be made when writing the constitutive relations in a derivative expansion, and each choice is called a hydrodynamic frame \cite{Kovtun:2012rj}. This concept is systematically explored in BDNK theory via the introduction of a new set of transport parameters that parameterizes the choice of hydrodynamic frame. The resulting general equations of motion can be causal, strongly hyperbolic, and stable for a subset of these transport parameters \cite{BemficaDisconziNoronha, Kovtun:2019hdm, Bemfica:2019knx, Hoult:2020eho, Bemfica:2020zjp, Abboud:2023hos}. This implies that relativistic viscous phenomena can, in principle, be sensibly described using these generalized first-order theories. 

Over the past several decades, there has been significant effort to include stochastic fluctuations in relativistic systems \cite{Calzetta:1997aj, Kovtun:2003vj, Dunkel:2008ngc, Kapusta:2011gt, Kovtun:2011np, Kovtun:2012rj, Young:2013fka, Kumar:2013twa, Young:2014pka, Murase:2016rhl, Kapusta:2014dja, Akamatsu:2016llw, Akamatsu:2017rdu, Sakai:2017rfi, Stephanov:2017ghc, Martinez:2017jjf, Singh:2018dpk, Nahrgang:2018afz, Martinez:2018wia, Murase:2019cwc, An:2019osr, Martinez:2019bsn, Rajagopal:2019xwg, An:2020vri, An:2019csj, De:2020yyx, Sakai:2020pjw, Nahrgang:2020yxm, Dore:2020jye, Du:2020bxp, Calzetta:2020wzr, Torrieri:2020ezm, Dore:2021xqq, Petrosyan:2021lqi, De:2022tkb, Pradeep:2022mkf, An:2022jgc, Abbasi:2022rum, Chen:2022ryi, Kuroki:2023ebq}. These works have used a number of approaches, but much of the recent focus has been on the field theory techniques earlier used to study non-relativistic fluids. While some of this work has used the Martin-Siggia-Rose (MSR) approach \cite{Granese:2022igc, Abbasi:2022aao}, many of the developments have focused on the Schwinger-Keldysh (SK) action on the closed time path \cite{Grozdanov:2013dba, Kovtun:2014hpa, Harder:2015nxa, Crossley:2015evo, Sieberer:2015hba, Haehl:2015pja, Haehl:2015uoc, Haehl:2016pec, Jensen:2017kzi, Glorioso:2017fpd, Liu:2018kfw, Chen-Lin:2018kfl, Jensen:2018hse, deBoer:2018qqm, Haehl:2018lcu, Ghosh:2020lel, Jain:2020vgc, Jain:2020zhu, Pantelidou:2022ftm, Baggioli:2023tlc, Akyuz:2023lsm}. In the latter, actions are constructed as an EFT from the underlying quantum mechanical system using the dynamical Kubo-Martin-Schwinger (KMS) symmetry \cite{Crossley:2015evo, Kubo:1957mj, Martin:1959jp}. This symmetry ensures that the nonlinear generalization of the fluctuation-dissipation theorem holds \cite{Wang:1998wg}, providing the benefit that actions can be constructed to account for nonlinear fluctuations.

In a previous paper \cite{Mullins:2023tjg}, some of us developed a framework for studying thermal fluctuations that incorporated the recent developments concerning causality, stability, and dissipation in the relativistic regime described above. This work provided a generalization of the non-relativistic results of Fox and Uhlenbeck \cite{doi:10.1063/1.1693183}, using the so-called information current \cite{Gavassino:2021kjm}. The framework was then used to study the inclusion of thermal fluctuations in a number of hydrodynamic models, including the case of Israel-Stewart theory in a general hydrodynamic frame \cite{Noronha:2021syv} at zero chemical potential. 

In this paper, we show how the work of \cite{Mullins:2023tjg} can be applied to construct effective actions for stochastic relativistic hydrodynamic models. Using the standard closed time path formulation, we show that in first-order theories, imposing causality and stability leads to unwanted behavior of correlation functions and noise correlators. This appears because these effective theories are only hydrodynamically stable and entropy production is guaranteed to be non-negative only on-shell in the regime of validity of the first-order expansion. Using the MSR approach, we then construct an effective action in terms of the information current, such that the underlying dynamics are guaranteed to be stable off-shell due to the Gibbs stability criterion \cite{Gavassino:2021cli}. By understanding how this new effective action transforms under the relevant discrete symmetries (time reversal, parity, charge conjugation), we identify a new $\mathds{Z}_2$ symmetry, analogous to a KMS transformation, that leaves the action invariant. This is the covariant manifestation of the symmetry identified in \cite{Guo:2022ixk}, which is used here to formulate effective actions for hydrodynamic systems in the presence of non-conserved currents.

This paper is organized as follows. In Sec.\ \ref{U1_gaugetheory}, stochastic fluctuations of a conserved current are studied in first-order BDNK theory using the Schwinger-Keldysh formulation. Motivated by several subtleties that appear in first-order theories, a new formalism for describing hydrodynamic fluctuations in Israel-Stewart, built using the information current, is presented in Sec.\ \ref{Sec:thermo_stable_fluct}. The effective action for this new approach is derived in Sec.\ \ref{Sec:MSR_action}, while the application of this approach to diffusion is presented in Sec.\ \ref{U1_israel_stewart} and \ref{Sec:Green_fns}. In Sec.\ \ref{Sec:Comparison_SK}, a modified KMS symmetry involving time reversal and parity is derived explicitly from the effective action. This symmetry is also obtained from microscopic approaches in Sec.\ \ref{Sec:detailed_balance}. Examples of this symmetry are applied to the problem of diffusion in Sec.\ \ref{Sec:KMS_U1_thermo_stable}. Finally, for the sake of completeness, in Appendix \ref{SK_comparison}, we provide a review of the Schwinger-Keldysh theory for effective actions on the closed time path.

\emph{Notation}: We use natural units $\hbar=c=k_B=1$ and a 4-dimensional Minkowski spacetime metric $g_{\mu\nu}$ with a mostly plus signature. Greek indices
run from 0 to 3, lower-case Latin indices run from 1 to 3, and upper-case Latin indices run across the space of thermodynamic variables. The four-momentum is written as $k^{\mu} = (\omega, \mathbf{k})$. To simplify the notation, we shall denote $|\mathbf{k}|=k$ when convenient.

\section{Effective action for first-order theories: relativistic diffusion}
\label{U1_gaugetheory}

In order to illustrate the interplay between causality, stability, and the stochastic formulation of first-order hydrodynamic theories, in this section we consider a simple example involving the dynamics of a conserved current associated with a global $U(1)$ symmetry. For simplicity, we assume that the conserved current $J^\mu$ is embedded in a medium with a constant background temperature $T>0$ and constant flow velocity given by a timelike future-pointing 4-vector $u^\mu$ (normalized such that $u^\mu u_\mu = -1$). We shall first discuss the case without noise and later implement stochastic fluctuations using the Schwinger-Keldysh approach. 

The conserved current $J^\mu$ can always be decomposed as
\be
J^\mu = \mathcal{N}u^\mu + \mathcal{J}^\mu
\label{define_J}
\ee
where $\mathcal{J}^\mu u_\mu=0$. At first order in derivatives, assuming a constant temperature and flow velocity, the constitutive relations at first order in derivatives are given by 
\bea
\mathcal{N} &=& n + \lambda\, T\, u^\alpha \partial_\alpha \left(\mu/T\right)\\ \nonumber
\mathcal{J}^\alpha &=&  - T \kappa\, \Delta^{\alpha\nu}\partial_\nu\left(\mu/T\right)  
\label{defineNandJ}
\eea
where $\mu$ is the chemical potential, $n = n(T,\mu)$ is the equilibrium density, $\kappa$ is the conductivity, $\Delta_{\mu\nu} = g_{\mu\nu}+u_\mu u_\nu$ is the rank-two projector transverse to $u^\mu$, and $\lambda$ is, in this case where $T$ and $u^\mu$ are constant, the single BDNK coefficient that parameterizes the hydrodynamic frame \cite{Kovtun:2019hdm}. At first order, this coefficient can be shifted by a hydrodynamic frame transformation involving the chemical potential \cite{Kovtun:2019hdm}, i.e, by a field redefinition of the form $ \mu \rightarrow  \mu + \alpha\, u^\lambda \partial_\lambda (\mu/T)$, so that   $\lambda \to \lambda + \chi \alpha$ where $\chi = (\partial n/\partial\mu)_T>0$ is the susceptibility and $\alpha$ is some coefficient. Within a first-order approach,  $\kappa$ is invariant under hydrodynamic frame transformations. 
Finally, we note that the dissipative corrections to the current vanish in global equilibrium.

The equation of motion for the chemical potential $\mu$ is simply given by the conservation of the current
\be
\partial_\mu J^\mu=0 \qquad \Longrightarrow \qquad  T u^\nu u^\alpha \partial_\nu \left(\lambda \partial_\alpha (\mu/T)\right) + T\chi u^\nu \partial_\nu (\mu/T) - T \Delta^{\nu\alpha}\partial_\nu \left(\kappa \partial_\alpha(\mu/T)\right)=0. 
\label{EOMcurrent}
\ee
It is straightforward to determine the conditions under which the equation of motion above is causal. This is done by recognizing that the principal part of the second-order differential operator acting on $\mu/T$ is given by  $T\left(\lambda\, u^\mu u^\nu  - \kappa \Delta^{\mu\nu}\right)\partial_\mu \partial_\nu$ \cite{ChoquetBruhatGRBook}. The corresponding characteristic polynomial is
\be
T\left(\lambda\, u^\mu u^\nu  - \kappa \Delta^{\mu\nu}\right)\phi_\mu \phi_\nu
\ee
where $\phi_\mu$ is the covector normal to the characteristic surface. Causality requires that the roots $\phi_\mu = (\phi_0(\phi_i),\phi_i)$ of the polynomial are real and $\phi_\mu \phi^\mu \geq 0$ \cite{Bemfica:2020zjp}. This occurs when 
\be
0 \leq \frac{\lambda}{\kappa} \leq 1.
\label{causality_current}
\ee
Thus, one can see that causality imposes that $\lambda$ cannot vanish and, in particular, assuming $\kappa >0$, $\lambda$ also cannot be negative. This result is in agreement with Refs.\ \cite{Hoult:2020eho,Abboud:2023hos}. Furthermore, if \eqref{causality_current} is satisfied, the retarded Green's function associated with this differential operator is guaranteed to vanish outside the future ``lightcone" defined by the characteristic speed $\kappa/\lambda$ \cite{WaldBookGR1984,ChoquetBruhatGRBook}, and linearized disturbances around the constant $\mu$ state can decay towards equilibrium, regardless of the Lorentz reference frame used. This is what hydrodynamic stability means in a relativistic system. In other words, causality ensures that a subluminal disturbance cannot be Lorentz transformed in a growing one, guaranteeing that the thermodynamic arrow of time points to the future in all Lorentz frames, not only in the rest frame \cite{Gavassino:2021owo}.

A standard Fourier analysis \cite{Hiscock:1985zz} of the equation of motion reveals that this system has a stable non-hydrodynamic mode with dispersion relation $\omega(\mathbf{k}\to 0) = - i\chi/\lambda$ in the local rest frame of the system where $u^\mu = (1,0,0,0)$. Then, we see that the BDNK coefficient regulates the dynamics by adding a stable non-hydrodynamic mode to the system that parameterizes our freedom to define the hydrodynamic variables out of equilibrium. Causality places an upper bound on the frequency of this non-hydrodynamic mode, which cannot exceed $1/D$, where the diffusion constant is $D = \kappa/\chi$. 

It is instructive at this point to remind the reader of the role played by BDNK coefficients in the second law of thermodynamics. Using the canonical construction for the entropy current \cite{Kovtun:2012rj}, one can use \eqref{EOMcurrent} to find 
\be
\partial_\alpha s^\alpha = T\left(\kappa \Delta^{\alpha\beta}-\lambda u^\alpha u^\beta\right)\partial_\alpha(\mu/T) \partial_\beta (\mu/T). 
\label{entropy_prod}
\ee
At this point, one would be tempted to say that the validity of the second law of thermodynamics requires that $\kappa\geq 0$ and $\lambda \leq 0$, which would be at odds with causality and stability. However, this reasoning is not correct, as explained in detail in \cite{Kovtun:2019hdm}. The entropy current is frame invariant in first-order hydrodynamics and, thus, one cannot use it to place constraints on the frame-dependent coefficient $\lambda$. In fact, the only physical requirement that $\partial_\mu s^\mu$ must obey is that it is non-negative in the regime of validity of the first-order theory \cite{Gavassino:2020ubn}. By implementing the derivative expansion for the on-shell quantities in \eqref{entropy_prod}, one can see that the term with $\lambda$ actually represents a third-order contribution to the entropy production, which cannot be accurately determined in the first-order approach. Thus, non-negative entropy production only implies that $\kappa \geq 0$, as expected. The coefficient $\lambda$ parameterizes our ignorance about the UV and, as such, in the first-order theory, it can be freely chosen to satisfy the causality and stability constraints. This choice should not affect predictions made using this theory (in the causal regime) if one stays in the regime of validity of first-order hydrodynamics. However, it should be noted that, since $\partial_\alpha s^\alpha$ can become negative at large gradients, the total entropy does not play the role of a rigorous Lyapunov functional \cite{Gavassino:2021cli}. Hence, we cannot use thermodynamic techniques to construct a standard information current \cite{Gavassino:2023odx}. Indeed, at present, there seems to be no way of associating a meaningful information current to first-order hydrodynamics \cite{Gavassino:2020ubn,Dore:2021xqq}.

Let us now investigate the stochastic formulation of this problem using the Schwinger-Keldysh framework \cite{Liu:2018kfw} (see Appendix \ref{SK_comparison} for a brief review). In particular, we follow the discussion and notation presented in \cite{Jain:2020zhu}, considering here a general hydrodynamic frame for the conserved current parametrized by $\lambda$ \cite{Hoult:2020eho}. The effective field theory is described using a phase field $\varphi_r$ and an associated stochastic noise field $\varphi_a$ \cite{Glorioso:2017fpd}, and background gauge fields $A_{r\mu}$ and $A_{a\mu}$. The effective theory is constructed using the gauge invariant Stueckelberg-like combination $B_{r,a\mu} = A_{r,a\mu}+\partial_{\mu}\varphi_{r,a}$. Up to leading order in derivatives, the SK effective action $\mathcal{S} = \int d^4 x\,\mathcal{L}_{\mathrm{SK}}$ is defined by the Lagrangian density given by
\begin{eqnarray} \label{U1_action_firstorder}
    \mathcal{L}_{\mathrm{SK}} & = &\, n u^{\nu} B_{a\nu}  + i T \left(\kappa \Delta^{\rho\nu}-\lambda u^\rho u^\nu\right)B_{a\rho}\left[B_{a\nu} + i \beta^\alpha F_{r \alpha\nu}+i\partial_\nu (\mu/T)\right],
\end{eqnarray}
where the chemical potential and temperature are defined using $\beta^\alpha = u^\alpha/T$ and $\mu/T = \beta^\alpha B_{r\alpha}$, and $F_{r\alpha\nu} = \partial_\alpha A_{r\nu}-\partial_\nu A_{r,\alpha}$. This action satisfies $\mathcal{S}[B_r,B_a=0]=0$, $\mathcal{S}[B_r,-B_a]=-\mathcal{S}^*[B_r,B_a]$, and the classical limit of KMS symmetry \cite{Jain:2020zhu}. The classical equation of motion in Eq.\ \eqref{EOMcurrent}, where $J^\mu$ is given by  \eqref{define_J} and \eqref{defineNandJ}, is obtained by varying this action with respect to $\varphi_a$, and setting $\varphi_a=0$, $A_{a\mu}=0$, and $\beta^\alpha F_{r\alpha \nu}=0$. 

It is convenient to write the Lagrangian density as follows 
\begin{equation} \label{U1_action_JB}
    \mathcal{L}_{\mathrm{SK}} = J^{\mu} B_{a\mu} + i\, T \left(  \kappa \Delta^{\mu\nu} - \lambda u^{\mu} u^{\nu} \right) B_{a\mu} B_{a\nu}.
\end{equation}
This form makes it clear that the usual constraint employed in Schwinger-Keldysh approaches, $\mathrm{Im} \,\mathcal{S}[B_r,B_a]\geq 0$, implies that $\kappa \geq 0$ and $\lambda \leq 0$. Since KMS symmetry dictates that these parameters are the same as the ones that appear in the classical equations of motion, requiring $\mathrm{Im} \,\mathcal{S}[B_r,B_a]\geq 0$ implies that the corresponding classical dynamics, taken at face value, would be acausal and unstable.  
Now, one may argue that, since $\lambda$ can be changed via a field redefinition (i.e., by changing the hydrodynamic frame), its value does not have an intrinsic meaning to first-order in gradients. Hence, it may seem natural that the inequality $\mathrm{Im} \,\mathcal{S}[B_r,B_a]\geq 0$ should only  constrain the frame invariant coefficient $\kappa$, leaving $\lambda$ unconstrained. However, now we shall show that, just like $\lambda > 0$ is required to guarantee that the classical dynamics is covariantly stable,  $\lambda \leq 0$ is required for the stochastic fluctuations to be stable. The complementarity of the above conditions will lead us to the striking conclusion that first-order fluctuating hydrodynamics cannot be covariantly stable, at least within the present approach. 

In order to show this, let us consider the Gaussian part of the effective Lagrangian, expanding around the equilibrium state with a constant chemical potential $\mu_0$ \cite{Jain:2020vgc}. For simplicity, let us also ignore the background gauge fields. Then, we can write the path integral weights explicitly:
\begin{eqnarray}
    |e^{iS}|&=&\exp\bigg[ T \int d^4 x \,  (-\kappa \Delta^{\nu\rho}+\lambda u^\nu u^\rho)\partial_\nu \varphi_a \partial_\rho \varphi_a \bigg] \\
    &=& \exp\bigg[ T \int \dfrac{d^4 k}{(2\pi)^4} \,  (-\kappa \Delta^{\nu\rho}+\lambda u^\nu u^\rho)k_\nu k_\rho  \, |\varphi_a(k^\mu)|^2 \bigg] ,
\end{eqnarray}
where in the second line we have expressed the integral in momentum space. Now, it is evident that, if $\lambda \leq 0$, the quantity $(-\kappa \Delta^{\nu\rho}+\lambda u^\nu u^\rho)k_\nu k_\rho$ is always negative, and the path integral weight is a regular Gaussian, where the equilibrium state $\varphi_a=0$ is the absolute maximum. However, if we force the dynamics to be causal (so that $\lambda >0$), then the Gaussian undergoes an inversion of convexity whenever $|\omega|>\sqrt{\kappa/\lambda}\, |k|$, i.e. for fluctuations outside the cone defined by the characteristic velocity. This implies that excitations of this kind, which are forbidden along classical solutions of the equations of motion, become favored when we turn on stochastic fluctuations since $(-\kappa \Delta^{\nu\rho}+\lambda u^\nu u^\rho)k_\nu k_\rho>0$. When this happens,  the equilibrium state ceases to be the most probable state, and it becomes unstable\footnote{We thank A.~Jain for discussions about this point.}. This is a manifestation of the breakdown of the maximum entropy principle in first-order hydrodynamics, which makes off-shell fluctuations entropically favored over the equilibrium state \cite{Gavassino:2020ubn}. Note that this is a problem also from a purely mathematical perspective because the path integral no longer converges since the weight grows like $\sim e^{+(...)\varphi_a^2}$.

Let us now compute the field-field correlators (for $\lambda \leq 0$). Under the assumptions mentioned above (i.e. expansion around an equilibrium state with constant chemical potential, and vanishing gauge fields),  the free part of the effective action reads
\begin{eqnarray}
    \mathcal{L}_{\mathrm{SK}} &=& -\varphi_a \left[\frac{\lambda}{\chi}  u^\nu u^\rho \partial_\nu\partial_\rho \delta n - D \Delta^{\nu\rho}\partial_\nu\partial_\rho \delta n +  u^\nu \partial_\nu \delta n  \right] \\ \nonumber &+& i T (\kappa \Delta^{\nu\rho}-\lambda u^\nu u^\rho)\partial_\nu \varphi_a \partial_\rho \varphi_a,
    \label{SKaction}
    \end{eqnarray}
where we used $\delta n = \chi \delta \mu$, with the coefficients evaluated in equilibrium. This leads to the following tree-level propagators, written in the local rest frame of the fluid for the sake of simplicity:
\bea
\langle \delta n(k^{\mu})\varphi_a(-k^{\mu})\rangle &=& \frac{1}{\omega + i \left(D k^2 - \frac{\lambda}{\chi}\omega^2\right)}\nonumber \\
\langle \varphi_a(k^{\mu}) \delta n(-k^{\mu})\rangle &=& \frac{-1}{\omega - i \left(D k^2 - \frac{\lambda}{\chi}\omega^2\right)}\nonumber \\
\langle \delta n(k^{\mu}) \delta n(-k^{\mu})\rangle &=& \frac{2T\left(D \chi k^2-\lambda \omega^2\right)}{\omega^2 + \left(\frac{\lambda}{\chi}\omega^2-D k^2\right)^2} = \frac{i T\chi}{\omega + i \left(D k^2 - \frac{\lambda}{\chi}\omega^2\right)}-\frac{i T\chi}{\omega -i \left(D k^2 - \frac{\lambda}{\chi}\omega^2\right)}\nonumber \\ 
\langle \varphi_a(k^{\mu}) \varphi_a(-k^{\mu})\rangle &=& 0.
\label{propagators}
\eea

In this formalism, the propagator $\langle \delta n(k^{\mu}) \varphi_a(-k^{\mu})\rangle$ should be retarded,  $\langle \varphi_a(k^{\mu}) \delta n(-k^{\mu})\rangle$ should be a purely advanced propagator, and $\langle \delta n(k^{\mu}) \delta n(-k^{\mu})\rangle$ should be a sum of retarded and advanced parts. However, this would be simultaneously true in all inertial frames only if the equation of motion were causal (which would require $\lambda>0$). Furthermore, only in causal systems, all inertial observers can agree on whether a disturbance caused by the noise is observed to cease and return the system to the equilibrium state or not \cite{Gavassino:2021owo}. If, instead, we modify a causal theory by performing some approximation in momentum space, then the propagators exit the lightcone \cite{Gavassino2023dispersion}, and one can always find two inertial observers who disagree on the chronological sequence of events within the disturbance. This leads to instability since, as pointed out in \cite{Gavassino:2021owo}, if the chronology of events is not the same for all observers, the cause of a given signal can be delayed and the system can spontaneously create a disturbance by taking entropy from the equilibrium state and reversing
the corresponding dissipative processes that would otherwise damp this perturbation.

Furthermore, we note that when $\lambda=0$ the standard arguments from Hiscock and Lindblom \cite{Hiscock:1985zz} hold and the number of modes obtained from the poles of the propagator changes from one, in the local rest frame, to two in the case of a general $u^\mu$ with nonzero 3-velocity, with the new mode being an unstable non-hydrodynamic mode.  
Similar arguments hold for $\langle \varphi_a(p^{\mu})\delta n(-k^{\mu}) \rangle$, with the new mode appearing on the wrong part of the complex $\omega$ plane, so that the function is not an advanced propagator anymore. 

On the other hand, let us imagine that we could continue \eqref{propagators} to positive $\lambda$, thereby enforcing causal propagation, see Eq.\ \eqref{causality_current}. Then, the average $\langle \delta n(k^{\mu}) \delta n(-k^{\mu})\rangle=\langle |\delta n(k^{\mu})|^2\rangle$, which should be non-negative by construction, becomes negative for $|\omega|>\sqrt{\kappa/\lambda}\, |k|$. This is the kind of mathematical inconsistency that we always meet when we try to extend a Gaussian average
\begin{eqnarray}
  \langle x^2 \rangle =  \dfrac{\int_\mathbb{R} e^{-ax^2} x^2 dx}{\int_\mathbb{R} e^{-ax^2}  dx} = \dfrac{1}{2a} 
\end{eqnarray}
to negative $a$, signaling that the integral is indeed not converging. This again shows that stochastic fluctuations are well defined (in the rest frame) only for acausal frames. Therefore, we have reached an impasse: whether we choose a causal hydrodynamic frame or not, the fluctuations are always ill-behaved according to some inertial observers. Of course, the pathological modes are of high frequency, and they formally fall outside the regime of applicability of the theory. Hence, one must be careful when using the propagators in Eq.\ \eqref{propagators}, especially when performing loops, to only remain in the regime of validity of the theory. The frame-dependent coefficient $\lambda$ defines the corresponding cutoff energy scale $\chi/\lambda$, defining the regime of applicability of the calculations.

This subtlety concerning the physical domain of propagators has interesting consequences for the noise correlator, as we show below. Let us first assume that $\lambda=0$. This is the case where the propagator $\langle \delta n \varphi_a\rangle$ is not retarded in all Lorentz frames, but $\langle \delta n \delta n\rangle$ is positive semi-definite for all $\omega$ and $k$, even outside of the regime of validity of the theory. In this case, one can go from the Schwinger-Keldysh path integral to another path integral with noise as follows \cite{Liu:2018kfw}. First, assume that one is in the local rest frame of the fluid and perform a Hubbard-Stratonovich transformation to a new variable $\xi^{i}$ such that the following term in the action can be written as 
\begin{equation}
i T \kappa \partial^j \varphi_a \partial_j \varphi_a = \frac{i}{4\kappa T}\xi^j \xi_j - \xi^j \partial_j\varphi_a . 
\end{equation}
After this transformation, the path integral then becomes
\begin{equation}
    Z \sim \int  \mathcal{D} \varphi_r \mathcal{D} \varphi_a\mathcal{D} \xi_{j} \exp \left\{ i \int d^4x \left[ \frac{i}{4 \kappa T} \xi_{j} \xi_{j} - \xi^{j} \partial_i \varphi_a + J^{\mu} \partial_\mu \varphi_a \right] \right\} .
\end{equation}
The auxiliary variable $\varphi_a$ now takes the role of a Lagrange multiplier, so it can be integrated out to obtain 
\begin{equation}
    Z \sim \int \mathcal{D}\varphi_r \mathcal{D} \xi_{\mu}\, \delta^{(4)} \left(\partial_\mu J^{\mu} - \partial_i\xi^{i} \right) \exp\left\{- \frac{1}{4\kappa T}\int d^4x\, \xi^{j}  \xi_{j}\right\}.
\end{equation}
This describes our conserved current with a stochastic source, 
\begin{equation}
\partial_{0} J^{0} + \partial_i (J^i-\xi^i)= 0,
\end{equation}
where the stochastic vector $\xi^{i}$ is sampled from a Gaussian distribution with zero mean and two-point correlator given by
\begin{equation}
    \langle \xi^{i}(x) \xi^{j}(x') \rangle =2 T \kappa\, \delta^{ij} \delta^{(4)}(x-x') .
    \label{noise_1}
\end{equation}
Note that this correlator is positive semi-definite.

Now, imagine that one wants to perform the same calculation with $\lambda=0$ but not in the local rest frame of the fluid, but with some arbitrary constant $u^\mu$. In this case,  one can decompose $\xi^\mu = - u^\mu (u_\nu \xi^\nu) + \xi^\mu_\perp$, where $\xi^\mu_\perp = \Delta^\mu_\nu \xi^\nu$, and the calculations proceed in the same way, leading to the noise correlator 
\begin{equation}
    \langle \xi_\perp^{\mu}(x) \xi_\perp^{\nu}(x') \rangle =2 T \kappa\, \Delta^{\mu\nu} \delta^{(4)}(x-x') .
    \label{noise_new}
\end{equation}

When $\lambda\neq 0$, we perform the Hubbard-Stratonovich transformation to find
\begin{equation}
    iT \left( \kappa \Delta^{\mu\nu} - \lambda  u^{\mu} u^{\nu} \right) \partial_\mu \varphi_a \partial_\nu \varphi_a = \frac{i}{4} \left(\frac{1}{\kappa T} \Delta^{\mu\nu} - \frac{1}{\lambda T} u^{\mu} u^{\nu} \right)\xi_{\mu} \xi_{\nu} - \xi^{\mu} \partial_\mu \varphi_a. 
\end{equation}
The path integral then becomes 
\begin{equation}
    Z \sim \int  \mathcal{D} \varphi_r \mathcal{D} \varphi_a\mathcal{D} \xi_{\mu} \exp \left\{ \int d^4x \left[ -\frac{1}{4} \left( \frac{1}{\kappa T} \Delta^{\mu\nu} - \frac{1}{\lambda T} u^{\mu} u^{\nu} \right) \xi_{\mu} \xi_{\nu} -i \xi^{\mu} \partial_\mu \varphi_a + iJ^{\mu} \partial_\mu \varphi_a \right] \right\} .
\end{equation}
Clearly, this only makes sense when $\lambda <0$ otherwise the path integral does not converge. Assuming $\lambda <0$ one can integrate out $\varphi_a$ to find
\begin{equation}
    Z \sim \int \mathcal{D}\varphi_r \mathcal{D} \xi_{\mu}\, \delta^{(4)} \left(\partial_\mu J^{\mu} - \partial_\mu\xi^{\mu} \right) \exp\left\{- \frac{1}{4}\int d^4x\, \xi_{\mu} \left( \frac{1}{\kappa T} \Delta^{\mu\nu} + \frac{1}{|\lambda T|} u^{\mu} u^{\nu} \right) \xi_{\nu}\right\},
\end{equation}
which describes our conserved current with a covariant stochastic source, 
\begin{equation}
\partial_{\mu} J^{\mu} = \partial_{\mu} \xi^{\mu} ,
\end{equation}
where the noise vector $\xi^{\mu}$ is sampled from a Gaussian distribution with zero mean and two-point correlator given by
\begin{equation}
    \langle \xi^{\mu}(x) \xi^{\nu}(x') \rangle = 2 T\left( \kappa \Delta^{\mu\nu} + |\lambda| u^{\mu} u^{\nu} \right) \delta^{(4)}(x-x') .
    \label{noise_2}
\end{equation}
In this case, the noise correlator is positive semi-definite. However, this occurs only when $\lambda \leq 0$, which precisely excludes the causal and stable region so the propagator $\langle \delta n \varphi_a\rangle$ is not retarded for all inertial observers. 

The discussion above can be easily generalized to consider the full BDNK equations \cite{BemficaDisconziNoronha,Kovtun:2019hdm,Bemfica:2019knx,Bemfica:2020zjp, Hoult:2020eho} including both the contributions coming from the energy-momentum tensor and the conserved current, and the same result will take place, namely: (i) in hydrodynamic frames where the dynamics is causal and stable, correlation functions computed using the Schwinger-Keldysh approach will only display the correct basic properties (e.g. $\langle \delta n \delta n\rangle$ being positive semi-definite) for $\omega$ and $k$ in the domain of validity of the theory; (ii) in the causal and stable regime, one cannot rewrite the Schwinger-Keldysh path integral in terms of a simpler path integral describing the conservation law in the presence of noise, this can  only be done in the acausal (hence, unstable  \cite{Gavassino:2021owo}) regime.

In the following sections, we change gears to consider the inclusion of stochastic fluctuations in \emph{second-order}, Israel-Stewart-type hydrodynamic theories where causality and stability in the linear regime can be demonstrated by properly defining a Lyapunov functional, which obeys the Gibbs stability criterion proposed in \cite{Gavassino:2021cli}, and ensures off-shell stability in all inertial reference frames. This leads to the definition of the so-called information current \cite{Gavassino:2021kjm}, which we use to determine the probability distribution of spontaneous fluctuations in the equilibrium state in a consistent manner, independently of the inertial reference frame. These elements are used to construct a theory for describing stochastic fluctuations in Israel-Stewart relativistic fluids.

\section{Effective action for Israel-Stewart theories}
\label{Sec:thermo_stable_fluct}

In \cite{Mullins:2023tjg}, a means of determining the noise correlators of Israel-Stewart-like relativistic hydrodynamic systems was presented. This approach relies on the so-called information current \cite{Gavassino:2021kjm}, defined as 
\begin{equation} \label{InformationCurrent_def}
    E^\mu = -\delta s^\mu - \alpha_I^* \delta J^{I\mu} ,
\end{equation}
where $s^\mu$ is the entropy current of the fluid, $\alpha^*_I$ are equilibrium constants (which refer to the environment), $J^{I\mu}$ are the different conserved currents associated with charges $Q^I$, and ``$\delta$" is an arbitrary finite perturbation of the equilibrium state. The information current tracks the net flow of information carried by perturbations around the equilibrium state. It is naturally related to the free energy, $\Omega$, of a relativistic system by
\begin{equation}
    \frac{\delta \Omega}{T} = \int d\Sigma\, n_{\mu} E^{\mu} ,
\end{equation}
where $\Sigma$ is an arbitrary spacelike hypersurface, and $n^{\mu}$ is the past-directed timelike unit normal to this hypersurface. This implies that the probability distribution for thermal fluctuations around equilibrium is given by 
\begin{eqnarray} \label{thermal_prob_dist}
    w[\delta \cvec{\phi}] \sim e^{- \int d\Sigma\, n_{\mu} E^{\mu}} ,
\end{eqnarray}
where $\delta \cvec{\phi}$ is a vector containing the perturbations of each thermodynamic variable around the equilibrium state.

The information current in Gibbs stable systems has the following properties \cite{Gavassino:2021kjm}:
\begin{enumerate}[i.]
    \item $E^{\mu} n_{\mu} \geq 0$ 
    for \emph{any} past-directed, timelike unit vector $n_{\mu}$.
    \item $E^{\mu} n_{\mu} = 0$ \emph{if and only if} the perturbation of each hydrodynamic variable, $\delta \cvec{\phi}$, is equal to zero. 
    \item $\partial_{\mu} E^{\mu} \leq 0$.
\end{enumerate}
Under these conditions, $\delta \Omega/T$ behaves as the Lyapunov functional used in the Gibbs stability analysis \cite{Gavassino:2021cli}. From the thermodynamic point of view, the Gibbs stability criterion ensures that any perturbation around equilibrium increases our knowledge about the microstates of the system. Furthermore, as shown in \cite{Gavassino:2021kjm}, the conditions above also guarantee causality in the linear regime. Finally, they also imply that the information current is unique \cite{Gavassino:2021kjm}. 

As mentioned in Sec.\ \ref{U1_gaugetheory}, working from Eq.\ \eqref{InformationCurrent_def} for first-order theories such as BDNK theory will yield an information current that does not have the properties mentioned above. This issue arises due to the fact that for such theories the equilibrium state will not be a maximum of the entropy \cite{Gavassino:2020ubn}. This does not affect the on-shell properties of such theories but introduces new subtleties to the off-shell formulation.

The approach we will use to determine the noise correlators is constructed entirely from the information current. It has been shown in  \cite{Gavassino:2023odx} that the equations of motion of any Gibbs stable relativistic linear system, as defined in \cite{Gavassino:2021kjm}, can be written in the form 
\begin{equation} \label{EoM_entropy_prod}
    \left( \cmat{E}^{\mu} \partial_{\mu} + \cmatgreek{\sigma} + \cmat{V}_{\mathrm{asym}} \right) \delta \cvec{\phi} = \cvec{\Xi} ,
\end{equation}
where $\sigma = \delta \cvec{\phi}^T \cmatgreek{\sigma} \delta \cvec{\phi}$ is the entropy production, and $\cmat{V}_{\mathrm{asym}}$ is the anti-symmetric part of $\cmat{V}$. The noise correlators in this approach take the form 
\begin{eqnarray} \label{Entropy_prod_noise}
    \langle \cvec{\Xi}(x) \cvec{\Xi}(x') \rangle = 2\cmatgreek{\sigma} \delta^{(4)}(x-x') .
\end{eqnarray}
This shows that the noise correlator scales with the entropy production, providing a very natural manifestation of the fluctuation-dissipation theorem. We will now determine the appropriate effective action to describe fluctuations in this system. 

\subsection{Martin-Siggia-Rose action}
\label{Sec:MSR_action}

It has been shown by MSR in \cite{Martin:1973zz} how stochastic differential equations can be written in terms of a path integral over an effective action. This path integral then defines a generating functional for correlation functions in the stochastic system. This approach has been used recently to study relativistic hydrodynamic fluctuations from kinetic theory in Ref.\  \cite{Granese:2022igc}. We will now show that the results presented in \cite{Mullins:2023tjg} can be formulated in the same way. 

Consider a system described by the equation of motion shown in Eq.\ \eqref{EoM_entropy_prod}. The average of any observable $\mathcal{O}$ that depends on the thermodynamic state of the system $\delta \cvec{\phi}$, can be written as the expectation value of a path integral 
\begin{equation}
    \langle \mathcal{O} \rangle = \left\langle \int \mathcal{D} \delta \cvec{\phi} \, \mathcal{O} \, \delta^{(4)} \left[ \left( \cmat{E}^{\mu} \partial_{\mu} + \cmatgreek{\sigma} + \cmat{V}_{\mathrm{asym}} \right) \delta \cvec{\phi} - \cvec{\Xi} \right] \right\rangle .
\end{equation}
The delta function can also be written as a path integral by introducing a new set of auxiliary variables, $\delta \bar{\cvec{\phi}}$, such that
\begin{equation}
\begin{split}
    \left\langle \mathcal{O} \right\rangle = & \left\langle \int \mathcal{D} \delta \cvec{\phi} \mathcal{D} \delta \bar{\cvec{\phi}} \, \mathcal{O} \exp \left\{ -i \int d^4x \, \delta \bar{\cvec{\phi}}^T \left[ (\cmat{E}^{\mu} \partial_{\mu} + \cmatgreek{\sigma} + \cmat{V}_{\mathrm{asym}}) \delta \cvec{\phi} - \cvec{\Xi} \right] \right\} \right\rangle .
\end{split}
\end{equation}
Since the noise is Gaussian, this expression can be written as
\begin{equation}
\begin{split}
    \left\langle \mathcal{O} \right\rangle = \int \mathcal{D} \delta \cvec{\phi} \mathcal{D} \delta \bar{\cvec{\phi}} \, \mathcal{O} \exp \Bigg[ & -i \int d^4x \, \delta \bar{\cvec{\phi}}^T(x) \left( \cmat{E}^{\mu} \partial_{\mu} + \cmatgreek{\sigma} + \cmat{V}_{\mathrm{asym}} \right) \delta \cvec{\phi}(x) \, + \\
    & - \frac{1}{2} \int d^4x \, d^4x' \, \delta \bar{\cvec{\phi}}^T(x) \langle \cvec{\Xi}(x) \cvec{\Xi}^T(x') \rangle \delta \bar{\cvec{\phi}}(x') \Bigg] .
\end{split}
\end{equation}
The expectation value of a quantity constructed using thermodynamic fields can thus be written as a path integral of the form 
\begin{equation}
    \left\langle \mathcal{O} \right\rangle = \int \mathcal{D} \delta \cvec{\phi} \mathcal{D} \delta \bar{\cvec{\phi}} \,\mathcal{O} \,e^{i S_{\mathrm{eff}}[\delta \cvec{\phi}, \delta \bar{\cvec{\phi}}]} ,
\end{equation}
where the effective action is given by 
\begin{equation} \label{MSR_action}
\begin{split}
    S_{\mathrm{eff}}[\delta \cvec{\phi}, \delta \bar{\cvec{\phi}}] = & -\int d^4x \,\delta \Bar{\cvec{\phi}}^T \left( \cmat{E}^{\mu} \partial_{\mu} + \cmatgreek{\sigma} + \tilde{\cmat{V}}_{\mathrm{asym}} \right) \delta \cvec{\phi} + i \int d^4x \,\delta \Bar{\cvec{\phi}}^T \,\cmatgreek{\sigma}\, \delta \bar{\cvec{\phi}} .
\end{split}
\end{equation}
Here, we have inserted the noise correlator found in \cite{Mullins:2023tjg}. We note that this action was constructed in \cite{Gavassino:2022roi} in the absence of stochastic fluctuations. A similar result can be obtained using other equivalent approaches to determine the fluctuations. For systems with no asymmetric term, $\tilde{\cmat{V}}_{\mathrm{asym}} = 0$, as is the case for most hydrodynamic models, the action is determined solely from thermodynamics.

The action in \eqref{MSR_action} describes the fluctuations of a multitude of relativistic systems. While here we will exclusively focus on applications in the context of Israel-Stewart theory, our results can be used to determine the properties of fluctuations in all of the different universality classes discussed in \cite{Gavassino:2023odx,Gavassino:2023qwl}, which categorize a myriad of physical systems ranging from heat conducting materials to relativistic superfluids and supersolids. 

\subsection{Israel-Stewart diffusion}
\label{U1_israel_stewart}

In Sec.\ \ref{U1_gaugetheory}, the thermodynamic fluctuations of a conserved current associated with a global $U(1)$ symmetry were discussed using first-order BDNK theory \cite{BemficaDisconziNoronha, Kovtun:2019hdm, Bemfica:2019knx,Bemfica:2020zjp,Hoult:2020eho}. Using the effective action constructed from the Schwinger-Keldysh framework \cite{Liu:2018kfw}, it was found that, if one imposes causality and stability, the correlation functions only display the expected physical properties for $\omega$ and $k$ in the hydrodynamic regime. We now discuss how the new approach presented here can be used to determine the stochastic fluctuations in a hydrodynamic theory of a $U(1)$ conserved current, in a general hydrodynamic frame, that is Gibbs stable according to the Gibbs stability criterion \cite{Gavassino:2021cli}.

To construct this type of Gibbs stable hydrodynamic description of a conserved current, we follow the formulation of Israel-Stewart theory \cite{MIS-6} in a general hydrodynamic frame worked out in \cite{Noronha:2021syv}. We again consider a conserved current of the form 
\begin{equation}
    J^{\mu} = (n + \mathcal{N}) u^{\mu} + \mathcal{J}^{\mu} ,
\end{equation}
where now not only $n$, but also  $\mathcal{N}$ and $\mathcal{J}^{\mu}$ are dynamical variables of the theory. The entropy current can be written up to second order in the degrees of freedom as,
\begin{equation}
    s^{\mu} = \left( s - \frac{\mu}{T} \mathcal{N} \right) u^{\mu} - \frac{\mu}{T} \mathcal{J}^{\mu} - \frac{u^{\mu}}{2T} \left( \beta_N \mathcal{N}^2 + \beta_J \mathcal{J}^{\lambda} \mathcal{J}_{\lambda} \right) ,
\end{equation}
where $\beta_N, \beta_J$ are new second-order transport coefficients. The second law of thermodynamics implies that the entropy production, $\sigma = \partial_{\mu} s^{\mu}$ is non-negative, indicating that it should be written as a quadratic form,
\begin{equation}
    \sigma = \frac{1}{T} \left( \frac{\mathcal{N}^2}{\lambda} + \frac{\mathcal{J}^{\alpha} \mathcal{J}_{\alpha}}{\kappa} \right) , 
\end{equation}
where $\kappa$ is the conductivity coefficient and $\lambda$ is a new transport coefficient. The three new transport coefficients $\beta_N, \beta_J, \lambda$ parameterize the hydrodynamic frame. This entropy production is non-negative if $\lambda, \kappa > 0$. Ensuring that the entropy production has this form requires the introduction of the relaxation equations, 
\begin{equation}
    \frac{\mathcal{N}}{\lambda T} = -\frac{\beta_N}{T} u^{\alpha} \partial_{\alpha} \mathcal{N} - u^{\alpha} \partial_{\alpha} \left( \frac{\mu}{T} \right) ,
\end{equation}
\begin{equation}
    \frac{\mathcal{J}^{\mu}}{\kappa T} = -\frac{\beta_J}{T} \Delta^{\mu}_{\:\: \nu} u^{\lambda} \partial_{\lambda} \mathcal{J}^{\nu} - \Delta^{\mu\nu} \partial_{\nu} \left( \frac{\mu}{T} \right) .
\end{equation}
Following \cite{Gavassino:2021kjm}, the information current of this theory is then given by 
\begin{equation}
    \begin{split}
        E^{\mu} & = -\delta s^{\mu} - \frac{\mu}{T} \delta J^{\mu} \\
        & = \frac{u^{\mu}}{\chi T} \left( \frac{1}{2} \delta n^2 + \delta n \delta \mathcal{N} \right) + \frac{1}{\chi T} \delta n \delta \mathcal{J}^{\mu} + \frac{u^{\mu}}{2T} \left( \beta_N \delta \mathcal{N}^2 + \beta_J \delta \mathcal{J}^{\lambda} \delta \mathcal{J}_{\lambda} \right) .
    \end{split}
\end{equation}
We must now determine the conditions for this theory to be Gibbs stable following the criterion \cite{Gavassino:2021cli} so that the new theory of fluctuations developed in Sec.\ \ref{Sec:thermo_stable_fluct} can be applied.

To find the conditions, one needs to fulfill the three criteria mentioned before. Property $\partial_{\mu} E^{\mu} \leq 0$ is simple to obtain in this Israel-Stewart formulation since $\partial_{\mu} E^{\mu} = -\sigma$, which implies that $\lambda, \kappa > 0$, precisely the condition for the entropy production to be non-negative mentioned above. The other two necessary conditions are restrictions on $n_{\mu} E^{\mu}$, where $n^{\mu}$ is an arbitrary past-directed, timelike 4-vector. Any $n^{\mu}$ with these properties can be decomposed up to an arbitrary positive constant as 
\begin{equation}
    n^{\mu} = \gamma (-1, v_n^i) ,
\end{equation}
where $\gamma = (1 - v_n^2)^{-1/2}$ is the Lorentz factor, and $v_n^2 \equiv v_n^i v_{ni} < 1$. Leaving $n^{\mu}$ in this general  form, the local rest frame of the fluid can be taken so $u^{\mu} = (1, 0,0,0)$. Then, 
\begin{equation}
\begin{split}
    \frac{2\chi T}{\gamma} n_{\mu} E^{\mu} & = \chi \beta_N \left( \frac{1}{\chi \beta_N} \delta n + \delta \mathcal{N} \right)^2 + \chi \beta_J \left( \frac{v_n^i}{\chi \beta_J} \delta n + \delta \mathcal{J}^i \right)^2 + \left( 1 - \frac{1}{\chi \beta_N} - \frac{v_n^2}{\chi \beta_J} \right) \delta n^2 .
\end{split}
\end{equation}
The first two terms are non-negative when $\beta_J, \beta_N \geq 0$; however, it is possible for them to be zero in a non-equilibrium state (for example $\chi \beta_J \delta \mathcal{J}^i = - v_n^i \delta n$). The last term must therefore be nonzero for the second Gibbs stability condition to hold. Finally, the last term must be non-negative to guarantee that the first condition holds. Hence, the system satisfies the Gibbs stability conditions for 
\begin{equation}
    \frac{1}{\chi \beta_N} + \frac{1}{\chi \beta_J} < 1 .
\end{equation}
It is thus possible to construct a stable system according to the Gibbs stability criterion using a $U(1)$ conserved current in this approach, so the new theory for fluctuations presented in \cite{Mullins:2023tjg} can be used. Note that this reduces to the theory of Sec.\ \ref{U1_gaugetheory} in the limit where $\beta_N, \beta_J \rightarrow 0$. 

The equations of motion of this theory are written in the form 
\begin{equation}
    \left( \cmat{E}^{\mu} \partial_{\mu} + \cmatgreek{\sigma} \right) \delta \cvec{\phi} = \cvec{\xi} ,
\end{equation}
where $\cvec{\xi}$ is a stochastic vector, $\delta \cvec{\phi} = \{ \delta n, \delta \mathcal{N}, \delta \mathcal{J}^{\mu} \}$, and 
\begin{equation}
    \cmat{E}^{\mu} = \frac{1}{\chi T} \begin{pmatrix}
        u^{\mu} & u^{\mu} & \Delta^{\mu\nu} \\
        u^{\mu} & \beta_N \chi u^{\mu} & 0 \\
        \Delta^{\mu}_{\:\: \rho} & 0 & \beta_J \chi u^{\mu} \Delta^{\nu}_{\:\:\rho}
    \end{pmatrix} ,
\end{equation}
\begin{equation}
    \cmatgreek{\sigma} = \frac{1}{T} \begin{pmatrix}
        0 & 0 & 0 \\
        0 & \frac{1}{\lambda} & 0 \\
        0 & 0 & \frac{1}{\kappa} \Delta^{\mu}_{\:\: \nu} 
    \end{pmatrix} .
\end{equation}
The noise correlator is then fixed by Eq.\ \eqref{Entropy_prod_noise} to be
\begin{equation}
    \langle \cvec{\xi}(x) \cvec{\xi}^T(x') \rangle = 2 \cmatgreek{\sigma} \delta^{(4)}(x-x') .
\end{equation}
This noise correlator is now positive semi-definite in the Gibbs stable, and hence causal, regime. Using the MSR approach, an action for this theory can be determined from Eq.\ \eqref{MSR_action}. Substituting the information current and entropy production above, one finds   
\begin{equation} \label{U1_action_general}
\begin{split}
    \mathcal{L}_{\mathrm{MSR}} = & - \frac{\delta \bar{n}}{\chi T} \left( u^{\mu} \partial_{\mu} \delta n + u^{\mu} \partial_{\mu} \delta \mathcal{N} + \partial_{\mu} \delta \mathcal{J}^{\mu} \right) - \frac{\delta \bar{\mathcal{N}}}{\chi T} \bigg( u^{\mu} \partial_{\mu} \delta n + \beta_N \chi u^{\mu} \partial_{\mu} \delta \mathcal{N} + \\
    & + \frac{\chi}{\lambda} \delta \mathcal{N} \bigg) - \frac{\delta \bar{\mathcal{J}}^{\mu}}{\chi T} \left( \partial_{\mu} \delta n + \beta_J \chi u^{\nu} \partial_{\nu} \delta \mathcal{J}_{\mu} + \frac{\chi}{\kappa} \delta \mathcal{J}_{\mu} \right) + \frac{i}{\lambda T} \delta \bar{\mathcal{N}}^2 + \frac{i}{\kappa T} \delta \bar{\mathcal{J}}^2 .
\end{split}
\end{equation}
The properties of this action will be discussed in Sec.\ \ref{Sec:KMS_U1_thermo_stable}, including how it behaves under the KMS symmetry described in Appendix \ref{Sec:dynamical_KMS}.

We can now determine the form of the symmetrized correlators from the equations of motion as follows. In momentum space, the equation of motion takes the form 
\begin{equation}
    \cmat{D}_J^{-1}(\omega,k)\, \delta \cvec{\phi}(\omega,k) = \cvec{\xi} ,
\end{equation}
with
\begin{equation}
    \cmat{D}_J^{-1}(\omega,k) =  -i\cmat{E}^0 \omega + i \cmat{E}^i k_i + \cmatgreek{\sigma} .
\end{equation}
The correlator of $\delta \cvec{\phi}$ is then given by 
\begin{equation}
    \langle \delta \cvec{\phi}(k^\mu) \,\delta \cvec{\phi}^T(-k^\mu) \rangle = \cmat{D}  (k^\mu) \cmatgreek{\sigma} \cmat{D}^{\dagger} ( k^\mu)  ,
\end{equation}
where we have substituted the correlator of $\cvec{\xi}$. The nonzero correlators of hydrodynamic variables thus take the form 
\begin{equation}
    \langle \delta n(k^{\mu}) \delta n(-k^{\mu}) \rangle = \frac{\chi^2 T \left[ \kappa \lambda \omega^2 \left( \lambda \beta_N^2 k^2 + \beta_J^2 \kappa \omega^2 \right) + \left( \kappa k^2 + \lambda \omega^2 \right) \right]}{\left| i \lambda \omega^2 + (i + \lambda \beta_N \omega) \left( \kappa k^2 - i \chi \omega \right) + \kappa \beta_J \omega^2 \left( -i\chi + \lambda \omega - \chi \lambda \beta_N \omega \right) \right|^2} ,
\end{equation}
\begin{equation}
    \langle \delta \mathcal{N}(k^{\mu}) \delta \mathcal{N}(-k^{\mu}) \rangle = \frac{T \lambda \left[ \kappa^2 k^4 + \left( \chi^2 + \kappa \lambda k^2 \right) \omega^2 + \chi \beta_J \kappa^2 \omega^2 \left( -k^2 + \chi \beta_J \omega^2 \right) \right]}{\left| i \lambda \omega^2 + (i + \lambda \beta_N \omega) \left( \kappa k^2 - i \chi \omega \right) + \kappa \beta_J \omega^2 \left( -i\chi + \lambda \omega - \chi \lambda \beta_N \omega \right) \right|^2}
\end{equation}
\begin{equation}
\begin{split}
    \langle \delta \mathcal{J}^{\mu}(k^{\alpha}) \delta \mathcal{J}^{\nu}(-k^{\alpha}) \rangle = \, & \frac{\kappa T \omega^2 \Delta^{\mu\nu}_{\alpha\beta} \left[ \chi^2 + \lambda \left( \kappa k^2 + \lambda \omega^2 \right) + \chi \beta_N \lambda^2 \omega^2 (\chi \beta_N - 2) \right]}{\left| i \lambda \omega^2 + (i + \lambda \beta_N \omega) \left( \kappa k^2 - i \chi \omega \right) + \kappa \beta_J \omega^2 \left( -i\chi + \lambda \omega - \chi \lambda \beta_N \omega \right) \right|^2} + \\
    & \,\,\,\,\, + \frac{\kappa T}{1 + \beta_J^2 \kappa^2 \omega^2} \Delta_{(k)}^{\mu\nu}
\end{split}
\end{equation}
\begin{equation}
    \langle \delta n(k^{\mu}) \delta \mathcal{N}(-k^{\mu}) \rangle = -\frac{\chi T \lambda \omega^2 \left[ \chi + \kappa \left( -\kappa \beta_J k^2 + \chi \kappa \beta_J^2 \omega^2 + \lambda \beta_N k^2 \right) \right]}{\left| i \lambda \omega^2 + (i + \lambda \beta_N \omega) \left( \kappa k^2 - i \chi \omega \right) + \kappa \beta_J \omega^2 \left( -i\chi + \lambda \omega - \chi \lambda \beta_N \omega \right) \right|^2}
\end{equation}
\begin{equation}
    \langle \delta n(k^{\nu}) \delta \mathcal{J}^{\mu}(-k^{\nu}) \rangle = \frac{\kappa \chi T \omega \left[ \chi + \lambda \omega^2 \left( \kappa \beta_J + \lambda \beta_N (\chi \beta_N - 1) \right) \right] \Delta^{\mu}_{\:\: \nu} k^{\nu}}{\left| i \lambda \omega^2 + (i + \lambda \beta_N \omega) \left( \kappa k^2 - i \chi \omega \right) + \kappa \beta_J \omega^2 \left( -i\chi + \lambda \omega - \chi \lambda \beta_N \omega \right) \right|^2}  ,
\end{equation}
where $\Delta^{\mu\nu}_{(k)} = \Delta^{\mu\nu} - \Delta^{\mu}_{\alpha} \Delta^{\nu}_{\beta} k^{\alpha} k^{\beta} / k^{\lambda} k_{\lambda}$ is the projector orthogonal to $k^{\mu}$ and $u^{\mu}$. These are the symmetrized correlators, but we would also like to obtain the retarded and advanced correlators. 

\subsection{Green's functions from the MSR action}
\label{Sec:Green_fns}

A key benefit of the Schwinger-Keldysh approach is that it allows for the determination of Green's functions, as described in Appendix \ref{SK_comparison}. To determine how the Green's functions can be obtained from the MSR action presented here, we must relate the sources to those that appear in Schwinger-Keldysh theory. We will work in the Eckart frame, for which there is no out-of-equilibrium scalar ($\mathcal{N} = 0$), for simplicity.

Generally, we expect that the sources should enter the effective action through a term of the form 
\begin{equation}
    S_{\mathrm{source}} = \int d^4x \left( \bar{A}_{\mu} J^{\mu} - B_{\mu} \bar{J}^{\mu} \right) ,
\end{equation}
where $\bar{A}_{\mu}$ is the Schwinger-Keldysh source that couples to the physical current, while $\bar{J}^{\mu}$ is some unknown auxiliary current and $B_{\mu}$ is its source. With these sources included, the equations of motion obtained by taking variations of the effective action with respect to $\delta \bar{\cvec{\phi}}$ are given by
\begin{equation}
    \partial_{\mu} J^{\mu} - \chi T u_{\mu} B^{\mu} = 0 ,
\end{equation}
\begin{equation}
    \partial^{\mu} \delta n + \frac{\beta_J \chi}{2} u^{\lambda} \partial_{\lambda} \delta \mathcal{J}^{\mu} + \frac{2\chi}{\kappa} \delta \mathcal{J}^{\mu} + \frac{i\chi}{\kappa} \delta \bar{\mathcal{J}}^{\mu} - \chi T B^{\mu} = 0 , 
\end{equation}
where we have defined $\delta \bar{J}^{\mu} = \delta \bar{n} u^{\mu} + \delta \bar{\mathcal{J}}^{\mu}$. These couple to the field tensor in the correct way if
\begin{equation}
    B^{\mu} = \beta_{\nu} F^{\nu\mu} ,
\end{equation}
where $F_{\mu\nu} = \partial_{\mu} A_{\nu} - \partial_{\nu} A_{\mu}$ is the field tensor. Inserting this in the source part of the effective action, 
\begin{equation}
\begin{split}
    S_{\mathrm{source}} & = \int d^4x \left( \bar{A}_{\mu} J^{\mu} - \beta_{\nu} F^{\nu}_{\:\:\: \mu} \bar{J}^{\mu} \right) \\
    & = \int d^4x \left( \bar{A}_{\mu} J^{\mu} + \beta_{\nu} A_{\mu} \partial^{\nu} \bar{J}^{\mu} \right) .
\end{split}
\end{equation}
In Schwinger-Keldysh theory, this last part of the action should have the form $-A_{\mu} J_a^{\mu}$, so we can identify
\begin{equation} \label{current_matching}
    J_a^{\mu} = \mathcal{L}_{\beta} \bar{J}^{\mu} . 
\end{equation}
A similar relation will be found in Sec.\ \ref{Sec:Comparison_SK} using the KMS symmetry. Here, however, we have found that the total action takes the form 
\begin{equation}
    S = S_{\mathrm{MSR}} + \int d^4x \left( \bar{A}_{\mu} J^{\mu} + A_{\mu} \mathcal{L}_{\beta} \bar{J}^{\mu} \right) ,
\end{equation}
from which the various Green's functions can be obtained by taking suitable variations. 

We now explicitly calculate the retarded Green's function for Israel-Stewart theory in the Eckart frame. This is simplest to do in matrix form, introducing $\cvec{\psi} = \{ \delta \cvec{\phi} , \delta \bar{\cvec{\phi}} \}$, and defining $\cvec{A}, \bar{\cvec{A}}$ by
\begin{equation}
    \cvec{A}^T \delta \bar{\cvec{\phi}} = A_{\mu} \bar{J}^{\mu} , \:\: \bar{\cvec{A}}^T \delta \cvec{\phi} = \bar{A}_{\mu} J^{\mu} .
\end{equation}
Working in Fourier space, the Lagrangian takes the form 
\begin{equation}
    \mathcal{L} = -\frac{1}{2} \cvec{\psi}^T \begin{pmatrix}
        0 & i\cmat{E}^{\mu} k_{\mu} + \cmatgreek{\sigma} \\
        -i \cmat{E}^{\mu} k_{\mu} + \cmatgreek{\sigma} & 2i \cmat{Q}
    \end{pmatrix} \cvec{\psi} - \begin{pmatrix}
        \bar{\cvec{A}} \\
        i\omega \cvec{A} / T
    \end{pmatrix}^T \cvec{\psi} .
\end{equation}
The resulting path integral is a Gaussian with a source, so we can complete the squares and integrate over the fields $\delta \cvec{\phi}, \delta \bar{\cvec{\phi}}$ to find that the generating functional is given by 
\begin{equation}
    W[A_{\mu}, \bar{A}_{\mu}] \sim i \begin{pmatrix}
        \bar{\cvec{A}} \\
        \frac{i\omega}{T} \cvec{A}
    \end{pmatrix}^{\dagger} \begin{pmatrix}
        0 & i\cmat{E}^{\mu} k_{\mu} + \cmatgreek{\sigma} \\
        -i \cmat{E}^{\mu} k_{\mu} + \cmatgreek{\sigma} & 2i \cmat{Q}
    \end{pmatrix}^{-1} \begin{pmatrix}
        \bar{\cvec{A}} \\
        \frac{i\omega}{T} \cvec{A}
    \end{pmatrix} .
\end{equation}
The Green's functions can then be identified from components of the inverse matrix, with suitable factors of $i\omega / T$ included. Note that this factor of $\omega / T$ is precisely the factor necessary for the fluctuation-dissipation theorem to hold. Here it appears due to the Lie derivative of Eq.\ \eqref{current_matching}.

Making the suitable variations, we find that the retarded correlators of the variables $\delta n, \delta \mathcal{J}^{\mu}$ take the form
\begin{equation}
    \cmat{G}_R = \langle \delta \cvec{\phi}^T \delta \bar{\cvec{\phi}} \rangle = \begin{pmatrix}
        \frac{\chi^2 \omega \left( \kappa \beta_J \omega + i \right)}{\kappa k^2 - i \chi \omega - \kappa \beta_J \chi \omega^2} & \frac{\kappa \chi \omega \Delta^{\mu}_{\nu} k^{\nu}}{\kappa k^2 - i \chi \omega - \kappa \beta_J \chi \omega^2} \\
        \frac{\kappa \chi \omega \Delta^{\mu}_{\nu} k_{\mu}}{\kappa k^2 - i \chi \omega - \kappa \beta_J \chi \omega^2} & \frac{\kappa \chi \omega^2 \Delta^{\mu\alpha} \Delta_{\nu}^{\beta} k_{\alpha} k_{\beta} / k^2}{\kappa k^2 - i \chi \omega - \kappa \beta_J \chi \omega^2} - \frac{\kappa \omega \Delta_{(k)\nu}^{\mu}}{i + \kappa \beta_J \omega} ,
    \end{pmatrix}
\end{equation}
where $\Delta_{(k)}^{\mu\nu}$ is the projector orthogonal to $k^{\mu}$ and $u^{\mu}$. Using these retarded correlators, the retarded Green's function for the current is found to be 
\begin{equation} \label{GR_Eckart}
\begin{split}
    G_R^{\mu\nu} = \, & \frac{\chi^2 \omega \left( \kappa \beta_J \omega + i \right) u^{\mu} u^{\nu} + \kappa \chi \omega ( \Delta^{\mu}_{\alpha} k_{\alpha} u^{\nu} + \Delta^{\nu}_{\alpha} k_{\alpha} u^{\mu}) + \kappa \chi \omega^2 \Delta^{\mu\alpha} \Delta^{\nu\beta} k_{\alpha} k_{\beta} / k^2}{\kappa k^2 - i \chi \omega - \kappa \beta_J \omega^2}  \\
    & - \frac{\kappa \omega}{i + \kappa \beta_J \omega} \Delta_{(k)}^{\mu\nu} .
\end{split}
\end{equation}
The advanced and symmetrized Green's functions can be obtained similarly, and the fluctuation-dissipation theorem is verified through a comparison of the retarded and symmetrized Green's functions. The approach used to obtain this Green's function can be employed to study systems with other conserved quantities by determining the appropriate source term in the action.

In the first-order limit, our calculations reduce to the standard result up to a contact term. To see this, consider the retarded $\delta n-\delta n$ correlator, 
\begin{equation}
    G_R^{nn} = \frac{\chi^2 \omega \left( \kappa \beta_J \omega + i \right)}{\kappa k^2 - i \chi \omega - \kappa \beta_J \chi \omega^2} .
\end{equation}
This can equivalently be written as 
\begin{equation} \label{GRnn_contact}
    G_R^{nn} = \frac{\chi \kappa k^2}{\kappa k^2 - i \chi \omega - \kappa \beta_J \chi \omega^2} - \chi ,
\end{equation}
A contact term, as appears here, is expected when determining Green's functions by variational techniques \cite{Kovtun:2012rj}. Taking the first-order limit requires taking $\beta_J \rightarrow 0$, so we are left with
\begin{equation}
    G_R^{nn} \rightarrow \frac{\chi \kappa k^2}{\kappa k^2 - i\chi \omega} = \frac{\chi D k^2}{Dk^2 - i \omega} ,
\end{equation}
where $D = \kappa / \chi$ and the contact term has been disregarded. This is now the standard result for the retarded Green's function of non-relativistic diffusion \cite{Kovtun:2012rj}. 

This contact term must be accounted for when using the retarded Green's function. For example, conservation of the current should imply that $k_{\mu} G_R^{\mu\nu} = 0$. However, performing this calculation with Eq.\ \eqref{GR_Eckart} we find that
\begin{equation}
    k_{\mu} G_R^{\mu\nu} = -\chi k_{\mu} u^{\mu} u^{\nu} .
\end{equation}
This apparent violation of the conservation law is solely due to the contact term found in Eq.\ \eqref{GRnn_contact}. Such terms frequently appear in momentum-space conservation laws for correlation functions for relativistic fluids \cite{Kovtun:2012rj}. By simply ignoring the contact term, the retarded Green's function can be written as
\begin{equation} \label{GR_Eckart_nocontact}
\begin{split}
    G_R^{\mu\nu} = \, & \frac{\chi \kappa k^2 u^{\mu} u^{\nu} + \kappa \chi \omega ( \Delta^{\mu}_{\alpha} k_{\alpha} u^{\nu} + \Delta^{\nu}_{\alpha} k_{\alpha} u^{\mu}) + \kappa \chi \omega^2 \Delta^{\mu\alpha} \Delta^{\nu\beta} k_{\alpha} k_{\beta} / k^2}{\kappa k^2 - i \chi \omega - \kappa \beta_J \omega^2}  \\
    & - \frac{\kappa \omega}{i + \kappa \beta_J \omega} \Delta_{(k)}^{\mu\nu} .
\end{split}
\end{equation}
This retarded Green's function will obey the standard conservation law-induced Ward identity.

\section{Discrete symmetries for Israel-Stewart theory}
\label{Sec:Comparison_SK}

The observant reader might find the derivation in Sec.\ \ref{Sec:MSR_action} familiar. It is precisely equivalent to the discussion involving Hubbard-Stratonovich transformations in Sec.\ \ref{U1_gaugetheory}, except in reverse. One might therefore expect that the action, Eq.\ \eqref{MSR_action}, is equivalent to the Schwinger-Keldysh action. However, there are two key differences. 

The first difference is that the Schwinger-Keldysh action is constructed to describe the dynamics of conserved quantities. For example, the action of Sec.\ \ref{U1_gaugetheory} describes the evolution of a conserved current $J^{\mu}$. 
For the MSR action constructed in this section, the dynamics can be written as 
\begin{equation}
    \partial_{\mu} (\cmat{E}^{\mu} \delta \cvec{\phi}) = -\cmatgreek{\sigma} - \tilde{\cmat{V}}_{\mathrm{asym}} . 
\end{equation}
The quantity $\cmat{E}^{\mu} \delta \cvec{\phi}$ can thus be thought of as a non-conserved current. To understand the interpretation of this current, recall
\begin{eqnarray}
    \frac{1}{2} \delta \cvec{\phi}^T \cmat{E}^{\mu} \delta \cvec{\phi} = E^{\mu} = -\delta s^{\mu} - \alpha_I^* \delta J^{I_{\mu}} .
\end{eqnarray}
Upon taking a derivative of $\cmat{E}^{\mu} \delta \cvec{\phi}$, there will be a nonzero term from the entropy current, assuming the system is dissipative, and potentially an antisymmetric term from $\delta J^{I_{\mu}}$, depending on the theory. Assuming no antisymmetric term for simplicity, the non-conserved dynamics comes from the entropy production.

The results of this section can thus be thought of as an application of the Schwinger-Keldysh approach to second-order hydrodynamic theories. First-order theories, such as BDNK theory \cite{BemficaDisconziNoronha, Kovtun:2019hdm, Bemfica:2019knx,Bemfica:2020zjp}, are directly amenable to the current Schwinger-Keldysh formalism \cite{Liu:2018kfw} because their equations of motion stem from the conservation laws. On the other hand, Israel-Stewart-like theories have non-conserved elements of their dynamics baked into the equations of motion, so their Schwinger-Keldysh formulation will have to start from a different starting point. Our results, therefore, extend the idea of using KMS symmetry to the case of hydrodynamic systems in the presence of non-conserved currents.\footnote{A similar extension using a standard Schwinger-Keldysh approach is provided in \cite{JainKovtun:Placeholder} for Israel-Stewart theory. Where the approaches overlap, the results of that work agree with those presented herein.}

The second key difference between the Schwinger-Keldysh and MSR approaches is that, in the former description, the ``quantum'' fields $\delta\cvec{\phi}_a$ have well-defined physical meaning in terms of the closed time path contour, and known transformation properties under time reversal (see Appendix~\ref{Sec:dynamical_KMS}).  
In the MSR formalism, however, the auxiliary fields $\delta\bar{\cvec{\phi}}$ are introduced to enforce the equations of motion and have no deeper meaning a priori. Below we show that the action obtained from our information-current-driven approach obeys, with appropriate modifications, the standard properties expected from a Schwinger-Keldysh action, providing a dictionary between these two approaches.
 
To more directly compare, we need to determine whether our action satisfies the Schwinger-Keldysh constraints 
\begin{eqnarray}
    S[\delta \cvec{\phi}_r, \delta \cvec{\phi}_a = 0] = 0 , \:\: S[\delta \cvec{\phi}_r, -\delta \cvec{\phi}_a] = -S^*[\delta \cvec{\phi}_r, \delta \cvec{\phi}_a] , \\
    \mathrm{Im} \left( S[\delta \cvec{\phi}_r, \delta \cvec{\phi}_a] \right) \geq 0 , \\
    S[\delta \cvec{\phi}_r, \delta \cvec{\phi}_a] = S[ \Theta \delta \cvec{\phi}_r, \Theta \delta \cvec{\phi}_a + i \Theta \mathcal{L}_{\beta} \delta \cvec{\phi}_r] ,
    \label{eq:KMS0}
\end{eqnarray}
where $\Theta$ represents the transformation of $\delta\cvec{\phi}$ under a discrete symmetry $\bm\Theta$ of the problem which includes time reversal $T$, and $\mathcal{L}_{\beta}$ is the Lie derivative with respect to the timelike Killing vector of the system. The first two conditions are trivially satisfied for the action of Eq.\ \eqref{MSR_action} provided that $\delta\cvec{\phi}_a \to - \delta\cvec{\phi}_a$ when $\delta\bar{\cvec{\phi}}\to -\delta\bar{\cvec{\phi}}$. However, the final condition, known as the dynamical KMS symmetry, requires a nontrivial link between the two descriptions. 

The KMS symmetry of the effective action is a remnant of the time reversal invariance of the microscopic theory (see Appendix~\ref{Sec:dynamical_KMS} for details). 
While the symmetry of the microscopic action, $S_{\mathrm{micro}}[\Theta\delta\cvec{\phi}_r,\Theta\delta\cvec{\phi}_a] = S_{\mathrm{micro}}[\delta\cvec{\phi}_r,\delta\cvec{\phi}_a]$, is lost due to the coarse-graining in the macroscopic effective theory, it can be recovered by swapping the initial and final states in the construction of the effective action $S_{\mathrm{eff}} \to S_{\mathrm{eff}}^{\bm\Theta}$, so that $S_{\mathrm{eff}}^{\bm\Theta}[\Theta\delta\cvec{\phi}_r,\Theta\delta\cvec{\phi}_a] = S_{\mathrm{eff}}[\delta\cvec{\phi}_r,\delta\cvec{\phi}_a]$ is still a symmetry (up to boundary terms). 
In the Schwinger-Keldysh formalism, $S_{\mathrm{eff}}^{\bm\Theta}$ can be found by changing the closed time path contour so that the initial and final density matrices are interchanged, which has the same effect as appropriately transforming the variables $\delta\cvec\phi_r$ and $\delta\cvec\phi_a$. In the classical limit, this transformation acts only in the ``quantum'' fields $\delta\cvec\phi_a$, and is given by the classical KMS symmetry of Eq.~\eqref{eq:KMS0}.

Here, rather than expecting that the auxiliary variables defined from the MSR action, $\delta \bar{\cvec{\phi}}$, transform in the same way as the Schwinger-Keldysh variables $\delta\cvec\phi_a$, we obtain their transformation properties 
from the MSR action of Eq.\ \eqref{MSR_action} itself. 
We do so by observing how the MSR action transforms under the discrete symmetry $\bm\Theta$, 
and enforcing the principle of detailed balance. 
Our procedure is very similar to the one of Ref.~\cite{Guo:2022ixk}, but it implements detailed balance in a way that is independent of the choice of space-time foliation.

\subsection{Modified KMS symmetry for the MSR action from time reversal} 

Consider first the transformation  $\delta\cvec{\phi}\to\Theta\,\delta\cvec{\phi}$ alone. We take $\bm\Theta$ to include both time reversal and parity. Because $E^\mu=\frac{1}{2}\delta\cvec{\phi}^T\,\cmat{E}^\mu\delta\cvec{\phi}$ is the information current, it transforms as $E^\mu \to E^\mu$, and it follows that $\cmat{E}^\mu \Theta = \Theta^T\,\cmat{E}^\mu$. By also demanding that the classical equation of motion in the absence of dissipation  $\left(\cmat{E}^\mu\partial_\mu + \tilde{\cmat{V}}_{\mathrm{asym}}\right)\delta\cvec{\phi}=0$ is preserved under $\Theta$, we find that $\tilde{\cmat{V}}_{\mathrm{asym}}\, \Theta = -  \Theta^T\,\tilde{\cmat{V}}_{\mathrm{asym}}$. On the other hand, as $\cmatgreek{\sigma}$ is responsible for dissipation and the breakdown of the symmetry under $\bm\Theta$, the dissipative equations of motion must transform to  $\left(\cmat{E}^\mu\partial_\mu + \tilde{\cmat{V}}_{\mathrm{asym}} - \cmatgreek{\sigma}\right)\delta\cvec{\phi}=0$, which implies $\cmatgreek{\sigma}\, \Theta = \Theta^T\,\cmatgreek{\sigma}$. 

To find how the auxiliary fields $\delta \bar{\cvec{\phi}}$ transform, consider an effective Lagrangian of the form 
\begin{eqnarray}
    \mathcal{L}_{\mathrm{MSR}} = -\delta \bar{\cvec{\phi}}^T \left( \cmat{E}^{\mu} \partial_{\mu} + \cmatgreek{\sigma} \right) \delta \cvec{\phi} + i \delta \bar{\cvec{\phi}}^T \cmat{Q}\, \delta \bar{\cvec{\phi}} ,
\end{eqnarray}
where the noise correlator $2\, \cmat{Q}$ has been left arbitrary. 
Again, the application of $\bm\Theta$ should not affect the ideal term, but should invert the dissipative one. Hence, we find that $\bm\Theta$ takes $\delta \bar{\cvec{\phi}} \to {\bm\Theta} \delta\bar{\cvec{\phi}} = - \Theta\, \delta \bar{\cvec{\phi}}$, and thus $\delta \bar{\cvec{\phi}}$ is of $\bm\Theta$-parity opposite to that of $\delta{\cvec{\phi}}$, analogously to the $T$-parity of canonically conjugate momenta.  
The Lagrangian above thus transforms as 
\begin{eqnarray}
    \mathcal{L}_{\mathrm{MSR}} \stackrel{\bm\Theta}{\longrightarrow} -\delta \bar{\cvec{\phi}}^T \left( \cmat{E}^{\mu} \partial_{\mu} - \cmatgreek{\sigma} \right) \delta \cvec{\phi}
    + i \,\delta \bar{\cvec{\phi}}^T \cmat{Q}_\Theta \,\delta \bar{\cvec{\phi}} ,
\end{eqnarray}
where we denote $\cmat{Q}_\Theta \equiv \Theta^T \cmat{Q} \,\Theta$.

To find the analogue of the KMS symmetry for our MSR action, we look for a  transformation of the form 
\begin{align}
\label{eq:tentativeKMSsym}
&\delta \cvec{\phi} \rightarrow \Theta\, \delta \cvec{\phi} ,&  
&\delta \bar{\cvec{\phi}} \rightarrow - \Theta\, \delta \bar{\cvec{\phi}} + i \,\Theta\, \cmat{O}\, \delta \cvec{\phi} ,&
\end{align}
with $\cmat{O}$ some differential operator, such that the effective Lagrangian is invariant up to a total derivative.   
 Under this transformation, the Lagrangian becomes
\begin{align}
\begin{split}\label{eq:pseudoKMS_PT_derivation}
\mathcal{L}_{\mathrm{MSR}} \to \mathcal{L}_{\mathrm{KMS}} =& \mathcal{L}_{\mathrm{MSR}} 
+2\, \delta \bar{\cvec{\phi}}^T\left( \cmatgreek{\sigma} + \cmat{Q}_\Theta\,\cmat{O}\right)\delta\cvec{\phi}
+ i \,\delta \bar{\cvec{\phi}}^T\left(\cmat{Q}_\Theta - \cmat{Q}\right)\delta \bar{\cvec{\phi}} + \\
&-i\,\delta\cvec{\phi}^T\left(\cmatgreek{\sigma} +\cmat{O}^T\cmat{Q}_\Theta\right)\cmat{O}\,\delta\cvec{\phi}-i\,\delta\cvec{\phi}^T\cmat{O}\,\cmat{E}^\mu\partial_\mu \delta\cvec{\phi}.
\end{split}
\end{align}
Therefore, the  Lagrangian is invariant if $\cmat{Q} = \cmat{Q}_\Theta$ and $\cmat{O}^T\cmat{Q} = \cmat{Q}_\Theta\,\cmat{O} = - \cmatgreek{\sigma}$ and the last term is a total derivative. By employing the fluctuation-dissipation theorem, $\cmat{Q} = \cmatgreek{\sigma}$, one immediately finds that $\cmat{O} = -\cmat{I}$. This implies that the proper symmetry involving time reversal and parity is 
\begin{align}
    &\delta \cvec{\phi} \rightarrow \Theta \delta \cvec{\phi} ,& 
    & \delta \bar{\cvec{\phi}} \rightarrow - \Theta \delta \bar{\cvec{\phi}} - i \Theta \delta \cvec{\phi} .&
\end{align}
Here, we have derived this transformation by direct study of the MSR action, but it can also be obtained by invoking a physical principle, as discussed below.

\subsection{Detailed balance condition and relation to the modified KMS symmetry}
\label{Sec:detailed_balance}

The tentative transformation rule in Eq.~\eqref{eq:tentativeKMSsym} can be completed and fully understood by considering the principle of detailed balance and imposing the correct equilibrium probability distribution  \cite{Guo:2022ixk}.
A similar approach was also employed using Crooks theorem in \cite{Torrieri:2020ezm} by considering infinitesimal variations of the foliation.
The equilibrium probability distribution $w[\delta \cvec{\phi}]$ found in  \eqref{thermal_prob_dist} can be made stationary by imposing the following detailed balance condition\footnote{The extension of the detailed balance principle to a microscopic symmetry $\bm\Theta$ including parity is discussed in \cite{LIFSHITZ19811}.}:
\begin{equation} \label{detailed_balance}
    P[\delta \cvec{\phi}_f(x \in \Sigma_f) | \delta \cvec{\phi}_0(x \in \Sigma_0)] \,w[\delta \cvec{\phi}_0] = P[ \Theta \delta \cvec{\phi}_0(x \in \Sigma_0) | \Theta \delta \cvec{\phi}_f(x \in \Sigma_f)]\, w[\delta \cvec{\phi}_f] ,
\end{equation}
where $P[\delta \cvec{\phi}_f(x \in \Sigma_f) | \delta \cvec{\phi}_0(x \in \Sigma_0)]$ denotes the conditional probability distribution that the system is in state $\delta \cvec{\phi}_f$ for $x$ in some final hypersurface $\Sigma_f$ given that it was in state $\delta \cvec{\phi}_0$ for $x$ in some initial hypersurface $\Sigma_0$. 
Using Eq.\ \eqref{thermal_prob_dist}, the condition in Eq.~\eqref{detailed_balance} can be written as 
\begin{equation} \label{detailed_balance_2}
    P[\Theta \delta \cvec{\phi}_0(x \in \Sigma_0) | \Theta \delta \cvec{\phi}_f(x \in \Sigma_f)] = e^{- \int_{\Sigma_0}^{\Sigma_f} d^4x\, \delta \cvec{\phi}^T \cmat{E}^{\mu} \partial_{\mu} \delta \cvec{\phi}}\, P[\delta \cvec{\phi}_f(x \in \Sigma_f) | \delta \cvec{\phi}_0(x \in \Sigma_0)] .
\end{equation}
Expressing the conditional probability distribution as a path integral,
\begin{equation}
    P[\Theta \delta \cvec{\phi}_0(x \in \Sigma_0) | \Theta \delta \cvec{\phi}_f(x \in \Sigma_f)] = \int_{\Theta\delta \cvec{\phi}_f(x \in \Sigma_f)
    }
    ^{\Theta\delta \cvec{\phi}_0(x \in \Sigma_0) 
    } 
    \mathcal{D} \delta \cvec{\phi}\, \mathcal{D}  \delta \bar{\cvec{\phi}} \; e^{i \int d^4x \,\mathcal{L}} .
\end{equation}
We would like to make the bounds of this path integral match those that will appear on the right-hand-side of Eq.\ \eqref{detailed_balance_2} when both conditional probability distributions are written as path integrals. This can be done by flipping the bounds of this path integral and changing the integration variables as $\delta\cvec{\phi} \to {\bm\Theta}\delta\cvec{\phi}$,  $\delta\bar{\cvec{\phi}} \to {\bm\Theta}\delta\bar{\cvec{\phi}}$. This amounts to a full spacetime reversal, taking $\mathcal{L}[\delta\cvec{\phi},\delta\bar{\cvec{\phi}}] \to \mathcal{L}_{\mathrm{KMS}} [\delta\cvec{\phi},\delta\bar{\cvec{\phi}}] \equiv \mathcal{L}_{\bm\Theta}[{\bm\Theta}\delta\cvec{\phi},{\bm\Theta}\delta\bar{\cvec{\phi}}]$, where $\mathcal{L}_{\bm\Theta}$ is the Lagrangian transformed due to the swapping of the initial and final states:
\begin{equation}
    P[\Theta \delta \cvec{\phi}_0(x \in \Sigma_0) | \Theta \delta \cvec{\phi}_f(x \in \Sigma_f)] = \int_{\delta \cvec{\phi}_0(x \in \Sigma_0) 
    }^{\delta \cvec{\phi}_f(x \in \Sigma_f) 
    } \mathcal{D} \delta \cvec{\phi}\, \mathcal{D} \delta \bar{\cvec{\phi}} \;e^{i \int d^4x\, \mathcal{L}_{\mathrm{KMS}}} .
\end{equation}
Expressing both sides of Eq.\ \eqref{detailed_balance_2} as a path integral, we find that 
\begin{equation}
    \int_{\delta \cvec{\phi}_0(x \in \Sigma_0) 
    }^{\delta \cvec{\phi}_f(x \in \Sigma_f) 
    } \mathcal{D} \delta \cvec{\phi} \mathcal{D} \delta \bar{\cvec{\phi}} \;e^{i \int d^4x \,\mathcal{L}_{\mathrm{KMS}}} = \int_{\delta \cvec{\phi}_0(x \in \Sigma_0) 
    }^{\delta \cvec{\phi}_f(x \in \Sigma_f) 
    } \mathcal{D} \delta \cvec{\phi}\, \mathcal{D} \delta \bar{\cvec{\phi}} \;e^{i \int d^4x \left( \mathcal{L} + i \delta \cvec{\phi}^T \cmat{E}^{\mu} \partial_{\mu} \delta \cvec{\phi} \right)} .
\end{equation}
Since the boundary conditions of the path integrals have been made the same and are arbitrary, this equality is satisfied only when the integrands are equal, which implies that 
\begin{equation}
\label{eq:LThetaDB}
    \mathcal{L}_{\mathrm{KMS}} = \mathcal{L} + i \delta \cvec{\phi}^T \cmat{E}^{\mu} \partial_{\mu} \delta \cvec{\phi} .
\end{equation}
That is, for detailed balance to hold, the action of parity and time reversal on the effective action should shift the Lagrangian by a total derivative term that is determined from the stationary distribution of Eq.\ \eqref{thermal_prob_dist}. 
This shift is provided by the transformation rule of Eq.\ \eqref{eq:tentativeKMSsym}. 
Comparing Eq.~\eqref{eq:LThetaDB} to Eq.~\eqref{eq:pseudoKMS_PT_derivation}, we find that detailed balance  is achieved by taking $\cmat{O}=-\cmat{I}$, where \cmat{I} is the identity. We also find that $\cmat{Q} = \cmatgreek{\sigma}$, which is the fluctuation-dissipation relation recently found in \cite{Mullins:2023tjg}. 

Hence, the proper $\mathds{Z}_2$ symmetry involving time reversal invariance for this action is then
\begin{align} \label{reduced_KMS_transform}
    &\delta \cvec{\phi} \rightarrow \Theta \delta \cvec{\phi} ,& 
    & \delta \bar{\cvec{\phi}} \rightarrow - \Theta \delta \bar{\cvec{\phi}} - i \Theta \delta \cvec{\phi} .&
\end{align} 
This symmetry should be used for the MSR action in Eq.\ \eqref{MSR_action} in place of the previous form of KMS symmetry discussed in the case of first-order theories. It provides a guide for the construction of effective actions for hydrodynamic systems where the equations of motion do not solely stem from conservation laws. 

In the case of time-reversal even fields $\delta\cvec{\phi}$, the transformation rule in Eq.~\eqref{reduced_KMS_transform} could also be obtained following the procedure introduced in \cite{Guo:2022ixk}, by noting that the conjugate momenta corresponding to $\delta\cvec{\phi}$ is $\cvec{\pi}=-\cmat{E}^0\delta\bar{\cvec{\phi}}$ and taking the equilibrium distribution to be $\exp\left(-\Phi \right)$, with $\Phi = \frac{1}{2}\int d^3x\, \delta\cvec{\phi}^T \cmat{E}^0\delta\cvec{\phi}$. The corresponding symmetry transformation becomes $\cvec{\pi} \to -\cvec{\pi} + i\,\cvec{\mu}$, with $\cvec{\mu} \equiv - \partial \Phi/\partial\delta\cvec{\phi} = \cmat{E}^0 \delta\cvec{\phi}$, thus recovering Eq.~\eqref{reduced_KMS_transform}.

\subsection{Mapping to Schwinger-Keldysh variables}
\label{Sec:Mapping_SK}

From Eq.~\eqref{reduced_KMS_transform}, the standard Schwinger-Keldysh variables, $\delta \cvec{\phi}_a, \delta \cvec{\phi}_r$, for the MSR action can be derived. Since the Schwinger-Keldysh action is invariant when these transform as
\begin{eqnarray} \label{KMS_standard}
    \delta \cvec{\phi}_a \rightarrow \Theta \delta \cvec{\phi}_a + i \Theta \mathcal{L}_{\beta} \delta \cvec{\phi}_r , \:\: \delta \cvec{\phi}_r \rightarrow \Theta \delta \cvec{\phi}_r ,
\end{eqnarray}
the proper form of $\delta \cvec{\phi}_a, \delta \cvec{\phi}_r$ as a function of $\delta \bar{\cvec{\phi}}, \delta \cvec{\phi}$ should transform in this manner under Eq.\ \eqref{reduced_KMS_transform}. This can be realized by defining
\begin{eqnarray} \label{Bar_to_a}
    \delta \cvec{\phi}_r = \delta \cvec{\phi} , \:\: \delta \cvec{\phi}_a = \mathcal{L}_{\beta} \delta \bar{\cvec{\phi}} .
\end{eqnarray}
This relationship between these variables should not be too surprising as the information current picture can arise due to making an order reduction with respect to the physical evolution equation, such as in Israel-Stewart theory. This order reduction can be generated via integrating by parts, which contributes a factor of $\mathcal{L}_{\beta}$, as in Eq.\ \eqref{Bar_to_a}. While these variables can in principle be used to write an action that obeys the standard KMS symmetry, Eq.\ \eqref{reduced_KMS_transform} provides a much simpler way to determine if the action is KMS invariant. Note that the relationship between auxiliary variables recovers the correct form of the auxiliary current, Eq.\ \eqref{current_matching}, found in Sec.\ \ref{Sec:Green_fns}.

\subsection{Modified KMS symmetry for Israel-Stewart diffusion in a general frame}
\label{Sec:KMS_U1_thermo_stable}

We now consider the application of the modified KMS symmetry to Israel-Stewart diffusion in a general hydrodynamic frame. From the derivation above and Eq.\ \eqref{MSR_action}, we find that the MSR action for the Israel-Stewart theory of diffusion is given by 
\begin{equation}
\begin{split}
    \mathcal{L}_{\mathrm{MSR}} = & - \frac{\delta \bar{n}}{\chi T} \left( u^{\mu} \partial_{\mu} \delta n + u^{\mu} \partial_{\mu} \delta \mathcal{N} + \partial_{\mu} \delta \mathcal{J}^{\mu} \right) - \frac{\delta \bar{\mathcal{N}}}{\chi T} \bigg( u^{\mu} \partial_{\mu} \delta n + \beta_N \chi u^{\mu} \partial_{\mu} \delta \mathcal{N} + \\
    & + \frac{\chi}{\lambda} \delta \mathcal{N} \bigg) - \frac{\delta \bar{\mathcal{J}}^{\mu}}{\chi T} \left( \partial_{\mu} \delta n + \beta_J \chi u^{\nu} \partial_{\nu} \delta \mathcal{J}_{\mu} + \frac{\chi}{\kappa} \delta \mathcal{J}_{\mu} \right) + \frac{i}{\lambda T} \delta \bar{\mathcal{N}}^2 + \frac{i}{\kappa T} \delta \bar{\mathcal{J}}^2 .
\end{split}
\end{equation}
This action is invariant under the modified KMS transformation, Eq.\ \eqref{reduced_KMS_transform}. To verify this, we consider the transformation
\begin{eqnarray}
    \delta n \rightarrow \delta n , \:\: \delta \bar{n} \rightarrow - \delta \bar{n} - i \delta n ,
\end{eqnarray}
\begin{eqnarray}
    \delta \mathcal{N} \rightarrow \delta \mathcal{N} , \:\: \delta \bar{\mathcal{N}} \rightarrow - \delta \bar{\mathcal{N}} - i \delta \mathcal{N} ,
\end{eqnarray}
\begin{eqnarray}
    \delta \mathcal{J}^{\mu} \rightarrow \delta \mathcal{J}^{\mu} , \:\: \delta \bar{\mathcal{J}}^{\mu} \rightarrow - \delta \bar{\mathcal{J}}^{\mu} - i \delta \mathcal{J}^{\mu} .
\end{eqnarray}
Then, in the local rest frame, each term transforms to 
\begin{equation}
\begin{split}
    - \frac{\delta \bar{n}}{\chi T} \left( \partial_t \delta n + \partial_t \delta \mathcal{N} + \partial_i \delta \mathcal{J}^i \right) \rightarrow & \frac{\delta \bar{n}}{\chi T} \left( -\partial_t \delta n - \partial_t \delta \mathcal{N} - \partial_i \delta \mathcal{J}^i \right) + \\
    & - \frac{i \delta n}{\chi T} \left( \partial_t \delta \mathcal{N} + \partial_i \delta \mathcal{J}^i \right) ,
\end{split}
\end{equation}
\begin{equation}
\begin{split}
    - \frac{\delta \bar{\mathcal{N}}}{\chi T} \bigg( \partial_t \delta n + \beta_N \chi \partial_t \delta \mathcal{N} + \frac{\chi}{\lambda} \delta \mathcal{N} \bigg) \rightarrow & \frac{\delta \bar{\mathcal{N}}}{\chi T} \bigg( -\partial_t \delta n - \beta_N \chi \partial_t \delta \mathcal{N} + \frac{\chi}{\lambda} \delta \mathcal{N} \bigg) + \\
    & - \frac{i\delta \mathcal{N}}{\chi T} \bigg( \partial_t \delta n - \frac{\chi}{\lambda} \delta \mathcal{N} \bigg) ,
\end{split}
\end{equation}
\begin{equation}
\begin{split}
    - \frac{\delta \bar{\mathcal{J}}^i}{\chi T} \left( \partial_i \delta n + \beta_J \chi \partial_t \delta \mathcal{J}_i + \frac{\chi}{\kappa} \delta \mathcal{J}_i \right) \rightarrow & \frac{\delta \bar{\mathcal{J}}^i}{\chi T} \left( -\partial_i \delta n - \beta_J \chi \partial_t \delta \mathcal{J}_i + \frac{\chi}{\kappa} \delta \mathcal{J}_i \right) + \\
    & - \frac{i \delta \mathcal{J}^i}{\chi T} \left( \partial_i \delta n - \frac{\chi}{\kappa} \delta \mathcal{J}_i \right) ,
\end{split}
\end{equation}
\begin{eqnarray}
    \frac{i}{\lambda T} \delta \bar{\mathcal{N}}^2 \rightarrow \frac{i}{\lambda T} \left( \delta \bar{\mathcal{N}}^2 + 2i \delta \bar{\mathcal{N}} \delta \mathcal{N} - \delta \mathcal{N}^2 \right) ,
\end{eqnarray}
\begin{eqnarray}
    \frac{i}{\kappa T} \delta \bar{\mathcal{J}}^2 \rightarrow \frac{i}{\kappa T} \left( \delta \bar{\mathcal{J}}^2 + 2i \delta \bar{\mathcal{J}}_i \delta \mathcal{J}^i - \delta \mathcal{J}^2 \right) .
\end{eqnarray}
Combining each of these, we find that 
\begin{equation}
\begin{split}
    \mathcal{L}_{MSR} \rightarrow & - \frac{\delta \bar{n}}{\chi T} \left( u^{\mu} \partial_{\mu} \delta n + u^{\mu} \partial_{\mu} \delta \mathcal{N} + \partial_{\mu} \delta \mathcal{J}^{\mu} \right) - \frac{\delta \bar{\mathcal{N}}}{\chi T} \bigg( u^{\mu} \partial_{\mu} \delta n + \beta_N \chi u^{\mu} \partial_{\mu} \delta \mathcal{N} + \\
    & + \frac{\chi}{\lambda} \delta \mathcal{N} \bigg) - \frac{\delta \bar{\mathcal{J}}^{\mu}}{\chi T} \left( \partial_{\mu} \delta n + \beta_J \chi u^{\nu} \partial_{\nu} \delta \mathcal{J}_{\mu} + \frac{\chi}{\kappa} \delta \mathcal{J}_{\mu} \right) + \frac{i}{\lambda T} \delta \bar{\mathcal{N}}^2 + \frac{i}{\kappa T} \delta \bar{\mathcal{J}}^2 .
\end{split}
\end{equation}
Since the equation of motion for this conserved current cannot be written as a single second-order equation for the density $\delta n$, finding an action that is invariant under the standard KMS symmetry is difficult. In this form, however, the action can be easily determined from the information current and the entropy production, and it still transforms as expected under the appropriate modified KMS transformation.

\section{Conclusions}
\label{conclusions}

In this paper, we have studied how causality and stability manifest in the effective actions of fluctuating, relativistic, dissipative hydrodynamic systems. This is first explored through the example of diffusion in first-order BDNK theory, investigated using the Schwinger-Keldysh formalism. We showed that the correlation functions of hydrodynamic fluctuations only display the expected physical properties at small frequencies and wavenumber,
i.e., within the expected regime of validity of the first-order approach, when causality and stability conditions are imposed. However, the corresponding generating functional does not converge in this case. This issue arises due to the fact that the equilibrium state is not a maximum of the entropy in such theories, which in principle allows for fluctuations to grow without bound. These off-shell subtleties do not affect the on-shell properties of the theory, which remain well-defined. Our results indicate that there are still unresolved issues when considering hydrodynamic fluctuations in relativistic first-order theories (already in the linear regime).

This motivated us to consider the new theory of relativistic fluctuations developed in \cite{Mullins:2023tjg} for Gibbs stable systems with an information current. In these systems, which include Israel-Stewart theories as an example, causality and stability hold, and the equilibrium state is guaranteed to be the maximum of the entropy in a covariant manner. These properties follow from the information current, which tracks the net flow of information carried by perturbations around the equilibrium state. By constructing a theory of fluctuations from the information current, these desirable properties are built-in, whereas other methods can seemingly fluctuate unstable systems with no obvious issues, as discussed in Sec.\ \ref{U1_gaugetheory}. This class of systems is described by conservation laws and \emph{also} additional equations of motion that describe the relaxation process associated with dissipative fluxes. Therefore, their dynamics is not solely given by conservation laws. Such systems have not been explored much from the perspective of Schwinger-Keldysh effective field theory. By using thermodynamic arguments, we provide a versatile approach for describing the stochastic fluctuations of relativistic systems that does not rely on microscopic dynamics.

In order to describe the fluctuations of Israel-Stewart theories through an effective action, at the linearized level, we showed in this paper that the following simple recipe can be applied: 
\begin{enumerate}[i.]
    \item Construct the information current and entropy production, starting from the underlying symmetries and degrees of freedom as explained in \cite{Gavassino:2023odx,Gavassino:2023qwl}.
    \item Derive the conditions for Gibbs stability using \cite{Gavassino:2021cli}.
    \item Write the corresponding MSR effective action, Eq.\ \eqref{MSR_action}.
    \item For the calculation of Green's functions, introduce source terms of the form $\bar{\cvec{h}}^T \delta\cvec{\phi} + \cvec{h}^T\mathcal{L}_\beta \delta\bar{\cvec{\phi}}$, with sources 
    $\cvec{h}$ and $\bar{\cvec{h}}$, as exemplified in Sec.~\ref{Sec:Green_fns}. 
\end{enumerate}
Since a general formula for the action is provided by Eq.\ \eqref{MSR_action} that holds for any information current and entropy production, the on-shell physics dictates the form of the effective action. In \cite{Gavassino:2023odx, Gavassino:2023qwl} it is shown that the information current and entropy production can be determined using only knowledge of the tensor structure of the degrees of freedom, and which of these degrees of freedom are dissipative. This organizing principle can be extended to the results of this work, allowing for the straightforward construction of universality classes for fluctuating hydrodynamics, which may be hard to obtain using other approaches.

To compare this MSR action to the Schwinger-Keldysh action, we studied how it behaves under discrete symmetries. As expected, it was found that the MSR action is not invariant under the standard dynamical KMS symmetry of \cite{Liu:2018kfw}, since the standard transformation was only prescribed for conserved hydrodynamic fields. Rather, it was found that it obeys a new symmetry under time reversal and parity, which can be found from a direct examination of the action or using detailed balance \cite{Guo:2022ixk}. This new symmetry can be used to determine the noise distribution of fluctuating relativistic systems constructed from an information current. We expect that this symmetry will be important to guide future research concerning the inclusion of nonlinear effects from fluctuations. In that case, one must determine the information current and entropy production beyond quadratic order and use the modified KMS symmetry of Sec.\ \ref{Sec:Comparison_SK} to determine the corresponding off-shell terms, mirroring the procedure used in Schwinger-Keldysh theory \cite{Jain:2020zhu}. Such a construction would also give a nonlinear probability distribution for fluctuations that could potentially be used in other approaches for solving stochastic systems \cite{barabasi1995fractal}.

The properties of this new MSR action were examined through the example of diffusion in the Israel-Stewart approach.
In this example, the dynamics do not follow exclusively from the conservation law. We note that for such systems the Schwinger-Keldysh approach can in principle be applied, see \cite{JainKovtun:Placeholder}. The approach developed in this paper, based on the information current and its modified KMS symmetry, provides a simple framework that can be applied to any system with an information current satisfying the conditions of Sec.\ \ref{Sec:thermo_stable_fluct}. 

By implementing the modified KMS condition developed in Sec.~\ref{Sec:Comparison_SK}, one should be able to enforce the fluctuation-dissipation theorem at the full nonlinear level in extensions of our effective action beyond quadratic order. Moreover, the mapping to Schwinger-Keldysh variables presented in Sec.~\ref{Sec:Mapping_SK}  enables one to compute general $n$-point Green's functions, with the correct approach to equilibrium, via a similar approach to that of Sec.~\ref{Sec:Green_fns}. The construction of a nonlinear MSR effective theory will be important for the renormalization of transport coefficients \cite{Kovtun:2011np} and for investigating the renormalization-group flow of the  newfound universality classes discussed in  Refs.~\cite{Gavassino:2023odx, Gavassino:2023qwl}. 

Furthermore, the nonlinear extension of our effective theory construction should be relevant to investigate the dynamics of higher-order fluctuations  \cite{Mukherjee:2015swa,An:2019csj,An:2020vri,An:2022jgc}, which are phenomenologically relevant in the search for the QCD critical point \cite{Stephanov:2008qz,Athanasiou:2010kw,Mukherjee:2016kyu,Nahrgang:2018afz,Bluhm:2020mpc,Pihan:2022xcl,Almaalol:2022xwv,MUSES:2023hyz}.

\section*{Note Added}

During the final stages of this work's completion, a draft of this paper was shared with A.~Jain and P.~Kovtun, and a draft of their work \cite{JainKovtun:Placeholder} was provided to us. Both works involve the study of effective actions for Israel-Stewart-like systems. The paper \cite{JainKovtun:Placeholder} focused on the corresponding effective field theory formulation through the Schwinger-Keldysh approach, while in this paper we focused on enforcing off-shell causality and stability constraints with the information current. Despite these differences in motivation, there are a number of similarities between our works, and both papers were further progressed through our mutual discussions. Here, we briefly summarize the similarities and differences of these papers, as well as what was developed during our correspondence.

Under the standard Schwinger-Keldysh formulation, the effective action is constructed using the underlying symmetries of the system and the corresponding conserved quantities. In Israel-Stewart theory, however, conservation laws are insufficient to describe the dynamics on their own. In this work, we circumvented this issue by using the information current which is constructed from thermodynamic quantities in a fashion that is natural for Israel-Stewart theory. On the other hand, in Ref.\ \cite{JainKovtun:Placeholder}, extended irreversible thermodynamics is used to add new degrees-of-freedom to the first law of thermodynamics that generate the out-of-equilibrium terms of Israel-Stewart theory when the Schwinger-Keldysh action is derived. It was found during our correspondence that this approach is not sufficient to find a unique dynamical KMS symmetry, but rather the same action can be obtained from different models, leading to distinct realizations of the KMS symmetry. These are referred to as ``alternate" prescriptions in \cite{JainKovtun:Placeholder}, one of which recovers the modified KMS symmetry discussed in this work, in the linear case. By comparing the actions found in our work to those found in \cite{JainKovtun:Placeholder}, an additional symmetry of the action Eq.\ \eqref{U1_action_general} in the Landau frame was found that corresponds to the standard KMS symmetry. While this symmetry may seem preferable for comparison between these two approaches, it is more complicated to generalize when working from the information current.

In summary, we believe that the equivalence between our framework and the framework of \cite{JainKovtun:Placeholder} is a manifestation of the duality between thermodynamics and statistical mechanics. In fact, our approach is entirely ``macroscopic'', as it assigns to the macrostates a probability distribution in terms of their hydrodynamic free energy (through the formula $\mathcal{P}\propto e^{-\Omega/T}$). On the other hand, the approach of \cite{JainKovtun:Placeholder} aims to compute all quantum correlators within an effective field theory formalism, and it claims a more direct connection with microphysics. We consider the agreement of the two methodologies as a strong confirmation of the self-consistency of both and of fluctuating relativistic hydrodynamics as a whole.

\section*{Acknowledgements}
We thank A.~Jain and P.~Kovtun for sharing a draft of their work during the final stages of this work's completion, for the many insightful discussions about this topic, and for providing comments on our manuscript. 
We also thank M.~Kaminski, K.~Jensen, N.~Pinzani-Fokeeva, G.~Torrieri, and A.~Lucas for enlightening discussions about action principles for stochastic hydrodynamics. NM and JN are supported in part by the U.S. Department of Energy, Office of Science, Office for Nuclear Physics
under Award No. DE-SC0021301 and DE-SC0023861. MH and JN were partly supported by the National Science Foundation (NSF) within the framework of the MUSES collaboration, under grant number OAC-2103680. MH and JN thank KITP Santa Barbara for its hospitality during ``The Many Faces of Relativistic Fluid Dynamics" Program, where this work's last stages were completed. This research was partly supported by the National Science Foundation under Grant No. NSF PHY-1748958. LG is partially supported by Vanderbilt's Seeding Success Grant.

\appendix

\section{Schwinger-Keldysh approach}
\label{SK_comparison}

For the sake of completeness, in this appendix, we briefly review a few points concerning Schwinger-Keldysh effective field theory and its formulation on a closed time path (CTP) \cite{Schwinger:1960qe, Keldysh:1964ud, Liu:2018kfw}. 

\subsection{Path integrals in the closed time path}
The CTP allows for the study of time-dependent thermal processes. Let us start from the case of unitary evolution. 
Consider, for instance, the following 2-point function
 in the Heisenberg representation:
\begin{align}\label{eq:ex2point}
   \left\langle \hat \phi_B(t_2)\, \hat \phi_A(t_1) \right\rangle &= \tr \left\{ \unitary{0}{t_2} \,\hat \phi_B\, \unitary{t_2}{t_1}\, \hat \phi_A\, \unitary{t_1}{0}\,\hat\rho_0\right\}
   \,,
\end{align}
where $\hat\rho_0$ is the thermal-equilibrium density operator and $\unitary{t_f}{t_i}$ is the unitary evolution operator from time $t_i$ to time $t_f$. Regardless of time ordering, this thermal average can be computed using the following generating functional:
\begin{equation}\label{eq:ZCTP}
    e^{W[\cvec{h}_1, \cvec{h}_2]} = \tr\left\{\unitary{\mathcal{T}}{0}{\cvec{h}_1} \, \hat\rho_0\,\unitary{0}{\mathcal{T}}{\cvec{h}_2}\right\}
    = \tr\left\{ \unitary{0}{\mathcal{T}}{\cvec{h}_2} \, \unitary{\mathcal{T}}{0}{\cvec{h}_1} \, \hat\rho_0\right\}\,,
\end{equation}
where $\mathcal{T}>t_1,t_2$ is some time far in the future. In the equation above, we have introduced sources $\cvec{h}_1$ and $\cvec{h}_2$ into the evolution operators $\unitary{\mathcal{T}}{0}$ and $\unitary{0}{\mathcal{T}}$, respectively. Note that $\unitary{t_f}{t_i}{\cvec{h}_i} \neq \unitary{t_f{-}t_i}{0}{\cvec{h}_i}$, because the sources $\cvec{h}_i$ are time-dependent, thus breaking time-translation invariance. 

The functional derivative of $\unitary{\mathcal{T}}{0}{\cvec{h}}$ with respect to the source $h^A$ evaluated at time $t$ (with $\mathcal{T}\geq t \geq 0$) is defined as follows:
\begin{equation}\label{gandalf}
    \frac{\delta \unitary{\mathcal{T}}{0}{\cvec{h}} }{\delta h^A(t)} = \dfrac{d}{d\varepsilon}\bigg|_{\varepsilon=0}  \unitary{\mathcal{T}}{0}{\cvec{h}_{At\varepsilon}} \, ,
\end{equation}
where  $(\cvec{h}_{At\varepsilon})^B(\tau):=h^B(\tau)+\varepsilon \delta^B_A \delta(t-\tau)$. 
If $\eta>0$ is infinitesimal, we can write
\begin{equation}\label{saruman}
    \unitary{\mathcal{T}}{0}{\cvec{h}_{At\varepsilon}} = \unitary{\mathcal{T}}{t{+}\eta}{\cvec{h}_{At\varepsilon}} \, \unitary{t{+}\eta}{t{-}\eta}{\cvec{h}_{At\varepsilon}}  \, \unitary{t{-}\eta}{0}{\cvec{h}_{At\varepsilon}}  \, .
\end{equation}
However, since the operators $\unitary{t_f}{t_i}{\cvec{h}_{At\varepsilon}} $ are solutions of 
\begin{equation}
   \dfrac{d}{dt_f} \unitary{t_f}{t_i}{\cvec{h}_{At\varepsilon}}=-i\hat H_{\cvec{h}_{At\varepsilon}}(t_f) \, \unitary{t_f}{t_i}{\cvec{h}_{At\varepsilon}}\, ,  \quad \quad \unitary{t_i}{t_i}{\cvec{h}_{At\varepsilon}}=\mathbb{I} \, ,
\end{equation}
we have that (for $\eta \rightarrow 0^+$)
\begin{equation}\label{gollum}
    \begin{split}
\unitary{\mathcal{T}}{t{+}\eta}{\cvec{h}_{At\varepsilon}}  ={}& \unitary{\mathcal{T}}{t}{\cvec{h}} \, , \\
\unitary{t{+}\eta}{t{-}\eta}{\cvec{h}_{At\varepsilon}}  = {}& \mathbb{I} -i \int_{t{-}\eta}^{t{+}\eta} \hat H_{\cvec{h}_{At\varepsilon}}(\tau) d\tau +\mathcal{O}(\varepsilon^2)   \, , \\
\unitary{t{-}\eta}{0}{\cvec{h}_{At\varepsilon}} ={}& \unitary{t}{0}{\cvec{h}} \, . \\
    \end{split}
\end{equation}
Hence, plugging \eqref{saruman} and \eqref{gollum} into \eqref{gandalf},  we obtain
\begin{eqnarray}
    \frac{\delta \unitary{\mathcal{T}}{0}{\cvec{h}}} {\delta h^A(t)} =-i \,\unitary{\mathcal{T}}{t}{\cvec{h}}\, \dfrac{\partial \hat H_{\cvec{h}}(t)}{\partial h^A(t)}\, \unitary{t}{0}{\cvec{h}}  \, .
\end{eqnarray}
Assuming that the perturbation to the Hamiltonian has the form $h^A \hat \phi_A$, we obtain 
\begin{align}\label{gondor}
    &\frac{\delta \unitary{\mathcal{T}}{0}{\cvec{h}_1}}{\delta h_1^A(t)} = \dfrac{1}{i} \unitary{\mathcal{T}}{t}{\cvec{h}_1}\,\hat \phi_A\,\unitary{t}{0}{\cvec{h}_1}\,,&
&\frac{\delta \unitary{0}{\mathcal{T}}{\cvec{h}_2}}{\delta h_2^A(t)} = \dfrac{1}{i} \unitary{0}{t}{\cvec{h}_2}\,\hat \phi_A\,\unitary{t}{\mathcal{T}}{\cvec{h}_2}\,.&
\end{align}

The generating functional $W$ allows for the calculation of the connected part of Eq.~\eqref{eq:ex2point} both for $t_1<t_2$ and $t_1>t_2$, respectively:
\begin{align}
   \left\langle \mathcal{T}_+\left[\delta\hat \phi_B(t_2)\, \delta\hat \phi_A(t_1)\right] \right\rangle  
    &=\dfrac{1}{i^2} \frac{\delta^2 W[\cvec{h}_1, \cvec{h}_2] }{\delta h_1^A(t_1)\delta h_1^B(t_2)}\, \bigg|_{\substack{{h_1=0}\\{h_2=0}}}
   ,
\\
   \left\langle \mathcal{T}_-\left[\delta\hat \phi_B(t_2)\, \delta\hat \phi_A(t_1)\right] \right\rangle  
    &=\dfrac{1}{i^2} \frac{\delta^2 W[\cvec{h}_1, \cvec{h}_2] }{\delta h_2^A(t_1)\delta h_2^B(t_2)}\, \bigg|_{\substack{{h_1=0}\\{h_2=0}}}
   .
\end{align}
Further, because the operator $\unitary{0}{T}{\cvec{h}_2}$ is applied after $\unitary{T}{0}{\cvec{h}_1}$ in the rightmost side of Eq.~\eqref{eq:ZCTP}, one can compute the same two-point function, regardless of time ordering by performing variations with respect to $h_2^B(t_2)$ and $h_1^A(t_1)$.

Equation \eqref{eq:ZCTP} admits a path integral representation in the CTP contour with an action $S_{\textrm{CTP}}$ given by
\begin{equation}
    e^{W[\cvec{h}_1, \cvec{h}_2]} = \int \mathcal{D}\cvec{\phi}_1 \,\mathcal{D}\cvec{\phi}_2\, e^{i\,S_{\textrm{CTP}}[\cvec{\phi}_1,\cvec{\phi}_2; \,\cvec{h}_1,\cvec{h}_2]}\,. 
\end{equation}
One can observe that the branch ordered forward in time corresponds to evolution of \emph{ket} states in the density operator with $U_{\cvec{h}_1}(\mathcal{T},0)$. Conversely,  the branch ordered backward in time corresponds to evolution of \emph{bra} states in the density operator with $U_{\cvec{h}_2}(0,\mathcal{T})$. 

Finally, it is convenient to introduce symmetric and antisymmetric combinations of sources and fields in the two branches of the CTP contour:
\begin{align}
    &\cvec{\phi}_r \equiv \frac{1}{2}\left(\cvec{\phi}_1 + \cvec{\phi}_2\right)\,,&
    &\cvec{\phi}_a \equiv \cvec{\phi}_1 -\cvec{\phi}_2\,,&
    &\cvec{h}_r \equiv \frac{1}{2}\left(\cvec{h}_1 + \cvec{h}_2\right)\,,&
    &\cvec{h}_a \equiv \cvec{h}_1 -\cvec{h}_2\,,&
\end{align}
so that variations with respect to the sources yield the advanced, retarded, and symmetrized Green's functions, respectively:
\begin{align}
    & \dfrac{1}{i^2} \frac{\delta W[\cvec{h}_{r,a}]}{\delta \cvec{h}_r(x_1)\delta \cvec{h}_a(x_2)} = G_A(x_1,x_2)\,,&
    & \dfrac{1}{i^2}\frac{\delta W[\cvec{h}_{r,a}]}{\delta \cvec{h}_a(x_1)\delta \cvec{h}_r(x_2)} = G_R(x_1,x_2)\,,&
\end{align}
\begin{align}
    & \dfrac{1}{i^2}\frac{\delta W[\cvec{h}_{r,a}]}{\delta \cvec{h}_a(x_1)\delta \cvec{h}_a(x_2)} = G_S(x_1,x_2)\,.&
\end{align}

\subsection{Effective field theories in the CTP}

In principle, one could integrate out fast modes to obtain an effective action  for slow degrees of freedom, $S_{\textrm{eff}} [\cvec{\phi}_{r,a}^{\textrm{slow}};\, \cvec{h}_{r,a}^{\textrm{slow}}]$, in the CTP or Schwinger-Keldysh contour:
\begin{align}
    e^{W[\cvec{h}_{1,2}]} 
    &= \int \mathcal{D}\cvec{\phi}_r^{\textrm{slow}} \,\mathcal{D}\cvec{\phi}_a^{\textrm{slow}}\, \left(\int \mathcal{D}\cvec{\phi}_r^{\textrm{fast}} \,\mathcal{D}\cvec{\phi}_a^{\textrm{fast}}\, e^{i\,S_{\textrm{CTP}}[\cvec{\phi}^{\textrm{fast,slow}}_{r,a}; \,\cvec{h}_{r,a}]}\right)
\\
    &=\int \mathcal{D}\cvec{\phi}_r^{\textrm{slow}} \,\mathcal{D}\cvec{\phi}_a^{\textrm{slow}}\, e^{i\,S_{\textrm{eff}}[\cvec{\phi}^{\textrm{slow}}_{r,a};\, \cvec{h}_{r,a}]}\,.
\end{align}
Henceforth, we will drop the superscript in $\cvec{\phi}_{r,a}^{\textrm{slow}}$ and $\cvec{h}_{r,a}^{\textrm{slow}}$, so as to make the notation simpler.  The integration over fast degrees of freedom will be implicit whenever we work with $S_{\textrm{eff}}[\cvec{\phi}_{r,a};\, \cvec{h}_{r,a}]$. 

In practice,  $S_{\textrm{eff}}[\cvec{\phi}_{a,r};\, \cvec{h}_{a,r}]$ is often constructed in the spirit of an effective field theory. That is, by considering all the terms allowed by the relevant symmetries and physical constraints and then applying a truncation scheme to obtain a finite number of couplings. 
Besides constraints from microscopic theories, effective actions in the CTP must satisfy three requirements, for consistency. Because $\hat\rho_0$ is Hermitian, taking the complex conjugate of Eq.~\eqref{eq:ZCTP} exchanges the two branches of the CTP. Hence, $-S_{\textrm{eff}}^*[\cvec{\phi}_r,\cvec{\phi}_a;\,\cvec{h}_r,\cvec{h}_a] = S_{\textrm{eff}}[\cvec{\phi}_r,-\cvec{\phi}_a;\,\cvec{h}_r,-\cvec{h}_a] $. For the path integral to be well defined, the stability condition $\operatorname{Im}
\, S_{\textrm{eff}}[\cvec{\phi}_r,\cvec{\phi}_a;\,\cvec{h}_r,\cvec{h}_a] \geq 0$ must be satisfied, which can be proved from  Eq.~\eqref{eq:ZCTP} as a Cauchy–Schwarz inequality. 
Finally, because any unitary operation preserves the trace of the density operator $\tr\{U \,\hat\rho_0 \,U^\dagger\}=1$, one finds $S_{\textrm{eff}}[\cvec{\phi}_a=0;\,\cvec{h}_a=0]=0$.

\subsection{Time reversal and Kubo-Martin-Schwinger symmetry}
\label{Sec:dynamical_KMS}

Because of the integration over fast modes, the evolution of $\cvec{\phi}_1$ and $\cvec{\phi}_2$ becomes intertwined, which reflects the fact that, in general, the evolution of the (reduced) density operator is no longer unitary (i.e., given by the von Neumann equation) after a partial trace. 
As a consequence of non-unitarity, time reversal symmetry is, in general, broken by dissipation. However, the symmetry properties of the microscopic theory under time reversal have important consequences for the effective action, which are discussed below.

Suppose the microscopic theory is invariant under a discrete $\mathcal{Z}_2$ symmetry operation $\bm\Theta$, which includes time reversal. We assume as well that the system starts at time $t=0$ from the equilibrium density matrix $\hat\rho_0 = e^{\beta_\alpha \hat P^\alpha}/Z$, with $\beta^\alpha = u^\alpha/T$, $u^\alpha$ the 4-velocity of the medium, and $T$ its temperature. Under $\bm\Theta$, the generating functional transforms according to
\begin{align}
    e^{W[\cvec{h}_1, \cvec{h}_2]} \to e^{W_{\bm\Theta}[{\bm\Theta}\,\cvec{h}_1, {\bm\Theta}\,\cvec{h}_2]} &= 
    \tr\left\{ 
    \unitary{0}{T}{{\bm\Theta}\,\cvec{h}_2(x^\alpha)}
    \;\hat\rho_0\;
    \unitary{T}{0}{{\bm\Theta}\,\cvec{h}_1(x^\alpha)}
    \right\}.
\end{align}
Using the invariance of the microscopic theory under this transformation, we find how $W[\cvec{h}_1, \cvec{h}_2]$ changes when ${\bm\Theta}$ is applied to the sources $\cvec{h}_{1,2}$:
\begin{align}
    e^{W[{\bm\Theta}\,\cvec{h}_1, {\bm\Theta}\,\cvec{h}_2]}
    = e^{W_{\bm\Theta}[\cvec{h}_1, \cvec{h}_2]}
    &= 
    \tr\left\{ 
    \unitary{0}{T}{\cvec{h}_2(x^\alpha)}
    \;\hat\rho_0\;
    \unitary{T}{0}{\cvec{h}_1(x^\alpha)}
    \right\}
    \\
    &=
    \tr\left\{ 
    \unitary{0}{T}{\cvec{h}_2(x^\alpha)}
    \;\hat\rho_0\; \hat\rho_0^{-1}\;
    \unitary{T}{0}{\cvec{h}_1(x^\alpha - i\,\beta^\alpha)}\;\hat\rho_0
    \right\}
    \\
    &=
    \tr\left\{ 
    \unitary{T}{0}{\cvec{h}_1(x^\alpha - i\,\beta^\alpha)}
    \;\hat\rho_0\; 
    \unitary{0}{T}{\cvec{h}_2(x^\alpha)}
    \right\}
    \\
    &= e^{W[\cvec{h}_1(x^\alpha+i\,\beta^\alpha), \cvec{h}_2(x^\alpha)]},
    \label{eq:KMS12}
\end{align}
where we use the fact that $\hat\rho_0 = e^{-i(i\,\beta_\alpha \hat P^\alpha)}/Z$, with $\hat P^\mu$ being the 4-momentum operator, performs a translation by an imaginary displacement $i\beta^\mu$. 

In Eq.~\eqref{eq:KMS12}, we observe that the original effective potential $W[\cvec{h}_{1,2}]$ is not recovered when the microscopic symmetry ${\bm\Theta}$ is applied to the sources $\cvec{h}_{1,2}$. 
The reason for that can be traced back to the fact that the time reversal exchanges the initial and final states. However, the equilibrium density matrix allows us to recover the original $W[\cvec{h}_{1,2}]$ by making the following transformation:
\begin{align}\label{eq:KMSsymh12}
    &\cvec{h}_1(x^\alpha) \to {\bm\Theta} \,\cvec{h}_1(x^\alpha + i\, \beta^\alpha),&
    &\cvec{h}_2(x^\alpha) \to {\bm\Theta} \,\cvec{h}_2(x^\alpha).&
\end{align}
Because of its relation to the usual KMS condition, the transformation above is called a KMS symmetry \cite{Crossley:2015evo, Sieberer:2015hba, Glorioso:2017fpd, Liu:2018kfw}. It guarantees that the correct equilibrium partition function is recovered, and imposes the fluctuation-dissipation theorem at the nonlinear level. 

Just as $W[\cvec{h}_{1,2}]$ corresponds to a trace over the final density matrix, the effective action $S_{\textrm{eff}}[\cvec{\phi}_{1,2};\,\cvec{h}_{1,2}]$ contains a partial trace over fast degrees of freedom. 
One can enforce Eq.~\eqref{eq:KMSsymh12} at the level of the effective action by imposing that $S_{\textrm{eff}}[\cvec{\phi}_{1,2};\,\cvec{h}_{1,2}]$ is invariant under Eq.~\eqref{eq:KMSsymh12} supplemented by the change
\begin{align}\label{eq:KMSsymphi12}
    &\cvec{\phi}_1(x^\alpha) \to \Theta \,\cvec{\phi}_1(x^\alpha + i\, \beta^\alpha),&
    &\cvec{\phi}_2(x^\alpha) \to \Theta \,\cvec{\phi}_2(x^\alpha),&
\end{align}
up to boundary terms --- that is, up to total derivative terms in the effective Lagrangian.

If powers of $\hbar$ are restored, the imaginary displacement promoted by $\hat \rho_0$ becomes  $i\hbar\beta^\mu$. In the classical limit $ \hbar\omega\ll  T$, where $\omega$ is the typical energy scale, Eqs.~\eqref{eq:KMSsymh12} and \eqref{eq:KMSsymphi12} become
\begin{align}
    &\cvec{h}_r(x^\alpha) \to \Theta \,\cvec{h}_r(x^\alpha),&
    &\cvec{h}_a(x^\alpha) \to \Theta \,\left(\cvec{h}_a(x^\alpha) + i\beta^\alpha\partial_\alpha \cvec{h}_r(x^\alpha)\right),&
    \label{eq:classKMSh}
    \\
    &\cvec{\phi}_r(x^\alpha) \to \Theta \,\cvec{\phi}_r(x^\alpha),&
    &\cvec{\phi}_a(x^\alpha) \to \Theta \,\left(\cvec{\phi}_a(x^\alpha) + i\beta^\alpha\partial_\alpha \cvec{\phi}_r(x^\alpha)\right),&
    \label{eq:classKMSphi}
\end{align}
where we have changed to $r$ and $a$ variables. Equations~\eqref{eq:classKMSh} and \eqref{eq:classKMSphi} define the KMS symmetry in the classical limit \cite{Crossley:2015evo, Sieberer:2015hba, Glorioso:2017fpd, Liu:2018kfw}.

\bibliography{references,References_Jorge}

\end{document}